\documentclass[AMA,STIX1COL]{WileyNJD-v2}

\usepackage{amsthm}
\usepackage{graphicx}
\usepackage{color}
\usepackage{graphics,graphicx,epsfig,subfigure}
\usepackage{latexsym}
\usepackage{mathrsfs}
\usepackage{amsmath}
\usepackage{float}
\usepackage{psfrag}
\usepackage{rotating}
\usepackage{longtable}
\usepackage{supertabular}
\usepackage[misc]{ifsym}
\usepackage{gensymb}
\usepackage{adjustbox}
\usepackage{hyperref,xcolor}

\articletype{Research Article}%

\received{}
\revised{}
\accepted{}

\begin{document}

\title{Probabilistic Seismic Hazard Assessment of Palghar
District, Maharashtra, India by considering spatially non-uniform
seismicity}

\author[1]{Suman Sinha}

\author[1]{S. Selvan}

\author[1]{Sachin Khupat}

\author[1]{Rizwan Ali}


\address[]{\orgdiv{Engineering Seismology Division}, \orgname{Central Water and Power Research Station}, \orgaddress{\state{Pune}, \country{India}}}

\corres{*Suman Sinha \email{suman.sinha.phys@gmail.com/\\suman.sinha@cwprs.gov.in}}


\abstract[Summary]{The Palghar district of Maharashtra has recently 
received attention because of frequent occurrences of earthquakes in its vicinity in the 
last few years since November, 2018 \cite{srin1}. The district area falls 
under seismic zone III, as per the 
seismic zonation map of India. As the recent earthquake activities have 
been preceded
by many major seismic events in the region, it necessitates to re-evaluate 
the level of seismic hazard of the area in a reliable and realistic way. 
With this aim in mind, the probabilistic seismic hazard map of Palghar
district with regard to $5\%$ damped Peak Ground Acceleration (PGA) and Pseudo 
Spectral Acceleration (PSA) at $0.2$ s and $1.0$ s for $10\%$ and $2\%$
probability of exceedance (PoE) in $50$ years at
engineering bedrock level is presented. The estimation of hazard are
performed in a finer grid resolution of 
0.02$\degree$ $\times$ 0.02$\degree$ and takes into consideration the
non-uniform distribution of earthquake probability within a seismic source zone 
(SSZ) and data-driven selection of appropriate Ground Motion Prediction
Equations (GMPEs). The spatial variation of the hazard maps,
reflected in he final results, demonstrates notable improvements
over the earlier studies.}

\keywords{Suitability of GMPEs, Smooth gridded seismicity, 
Probabilistic seismic hazard, Log-likelihood}

\maketitle

\section{Introduction}
\label{intro}
\begin{figure}[!h]
\begin{center}
        \rotatebox{0}{\includegraphics[scale=0.6]{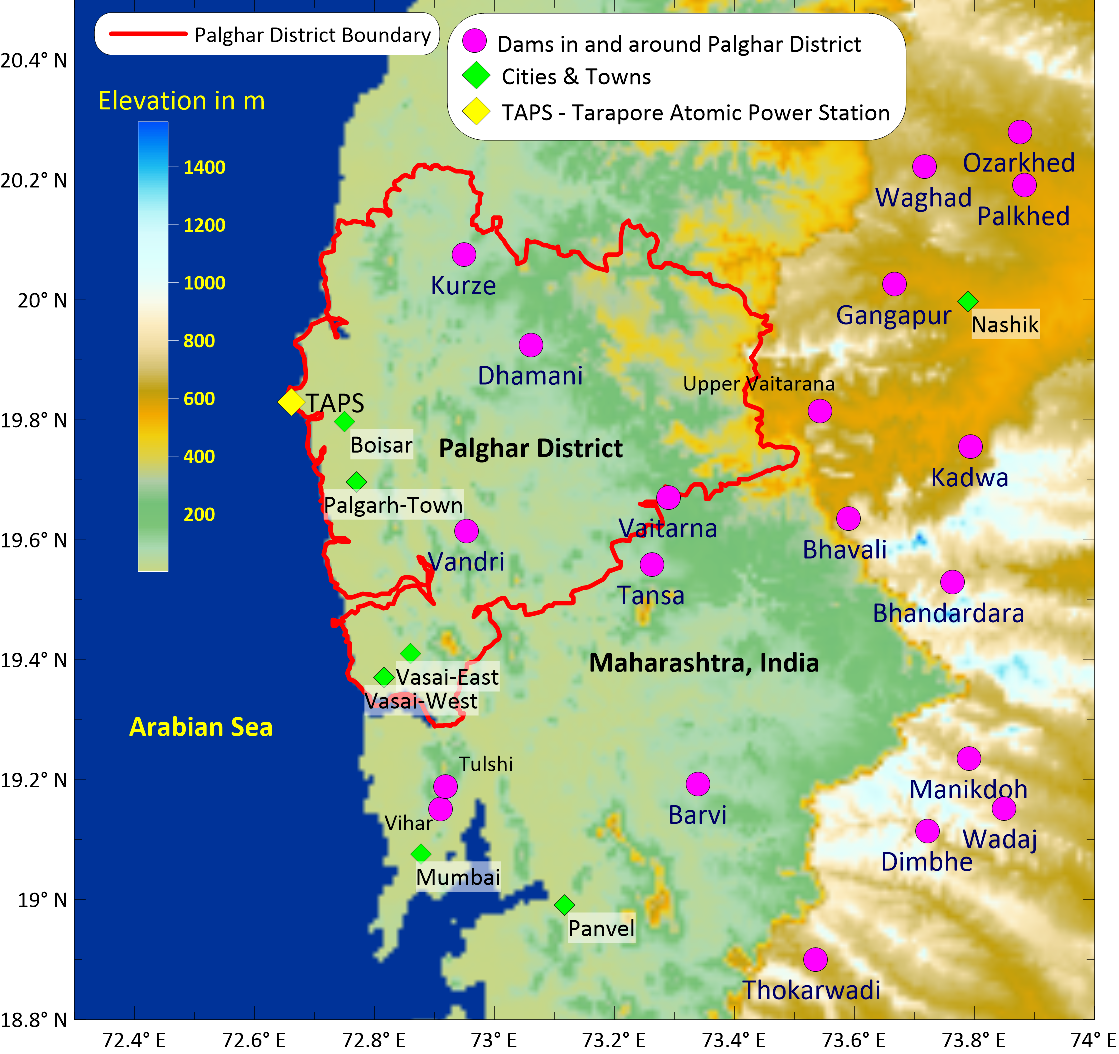}}
	\caption{The geographical boundary (approximate) of Palghar
        district and locations of the dams and Tarapore Atomic Power
        Station (TAPS).}
\label{ros}       
\end{center}
\end{figure}

Palghar district lies in the Konkan division of Maharashtra State
and Palghar town is the administrative capital of the Palghar district.
Palghar is considered to be a district of economical significance
in India and is an industrial hub, too. It is home to India's first 
atomic power plant, Tarapur Atomic Power Station (TAPS).
The busy Mumbai-Ahmedabad rail corridor passes through this district.
Moreover, a number of dams are housed in and around the Palghar district.
In the last four years, Palghar region witnessed an unusual frequency 
of earthquakes. Srinagesh \textit {et al.}\cite{srin1} stated that around 
4854 events
in the magnitude range 0.1 $M_L$ to 4.1 $M_L$ of focal depths ranging from 
4 km to 16 km during the period January 30, 2019 to August 31, 2019.
Within the Palghar district boundary, 34 events having magnitude
$M_L$ 3.0 to $M_L$ 4.1 have been reported by National Centre for
Seismology, 
Government of India during the period November, 2018 to October, 2019.
The largest ($M_L$ 4.1) among them took place on March 1, 2019
and eight earthquakes in the magnitude range $M_L$ 3.1 to 3.7
were recorded on February 1, 2019 at a span of 8 hours.
Although Palghar district falls under Peninsular shield which lies in the
Stable Continental Region (SCR) of the Indian Subcontinent,
the same has witnessed few major earthquakes in the past.
These include Bhuj earthquake ($M_W$ 7.6, 2001), Jabalpur earthquake 
($M_W$ 5.8, 1997), Latur earthquake ($M_W$ 6.1, 1993) and
Koyna earthquake ($M_W$ 6.4, 1967) which caused enormous loss of lives and
extensive damage to properties.
Therefore, re-evaluation of seismic safety to assess the vulnerability
of important structures against earthquakes is very much necessary for
this region. 
Map showing the boundary of Palghar district along with the
locations of important structures in its vicinity are shown in Figure 
\ref{ros}.

Seismic safety is generally evaluated in terms of seismic hazards
which are quantified by the determination of different
ground motion parameters.
The ground shaking associated with earthquakes plays the most vital role
\citep{kra1}. It inhabits in area prone to earthquakes with favourable seismo-tectonics 
and local geological site characteristics, causing significant damage 
to properties and lives. It necessitates estimating the terrain's seismic hazard 
realistically in order 
to derive the essential design criteria for building earthquake-resistant structures.
The sudden recent seismic activities, discussed earlier,
together with the past damaging earthquakes motivate
us to quantify seismic hazard in a realistic approach.

As per the seismic zoning map of India issued by Bureau of Indian Standards (BIS), it designates four classes of seismicity based on 
zone factor \citep{bis1}, which is double the zero-period acceleration (ZPA) of the 
design spectrum. The study area falls in Seismic Zone III with the zone 
factor $0.16$ as per BIS \cite{bis1}. 
But the main limitation of the Indian seismic zonation code, as presented 
in BIS \cite{bis1} is that it lacks probabilistic features \citep{khat1} and 
hence it isn't grounded in a thorough seismic hazard analysis. But given its solid 
scientific foundation \citep{kra1}, the probabilistic approach to seismic hazard 
calculation is thought to be more appropriate. Taking into account the randomness 
of earthquake occurrences in space, 
time, and magnitude, the ground motion, estimated using the PSHA method 
with a pre-defined confidence level, will not exceed at any time period 
due to any anticipated earthquake over a fixed period of time.
The mathematical expression of PSHA was formed by 
Cornell \cite{corn2} and McGuire \cite{mcg1}.

Some studies of PSHA of Peninsular India
\citep{gupta6,ashi1,ndma} have been carried out in the recent past.
These studies are quite silent about the importance of the ranking-based 
selection criteria of appropriate GMPEs. Although Scaria {\it{et al.}} \cite{gupta6}
had discussed the ranking of GMPEs, the authors finally chose an old
GMPE of Abrahamson and Silva \cite{abra1} with some modification of the amplitude term.
As the seismicity is, in general, diffused in nature in Peninsular shield, the
rates of occurrence of expected earthquakes in different magnitude ranges
are distributed non-uniformly over equally-spaced grids of interval
as per the smoothed
epicentral density of observed past earthquakes.
The approach is known as smooth-gridded seismicity model after
Frankel \citep{fran1}. On the contrary, the conventional uniform
seismicity model assumes same earthquake occurrence rates for
all the grid points inside a seismic source zone (SSZ), which
either underestimates or overestimates the earthquake occurrence rates,
depending upon the locations of past earthquakes. In this study, the 
grid interval is taken as
$0.02 \degree \times 0.02 \degree$.

The values of different ground motion parameters depend a lot on the
choice of the GMPEs. Therefore, selection of appropriate GMPEs for a
specific area is very crucial in any study on seismic hazard. The present
work has addressed this issue by carrying out a detailed quantitative
assessment for selection of suitable GMPEs.
The seismic hazard computations at higher resolution grids have been performed 
after assessing the suitability of various GMPEs against observed accelerograms 
using log-likelihood(LLH) score introduced by Scherbaum {\it et al.}\cite{sche2}. 

At engineering bedrock (which conforms to $V_{s30}$ value $\sim$ 760 m/s), 
the spatial distribution 
of PGA and $5\%$ damped PSA at specific 
time periods for $10\%$ and $2\%$ probability of exceedance (PoE) in $50$ years 
corresponding to return periods $475$ years, known as design basis earthquake (DBE)  
and $2475$ years, known as maximum credible earthquake
(MCE), respectively 
have been obtained \citep{ASCE,ICC}. As the periods 0.2 s and 1.0 s are 
commonly used as corner periods to create 
a smooth design spectrum for structural design \citep{nath4}, the PSA at these periods are 
estimated along with PGA. The resulting hazard maps show the 
spatial variations in seismic hazard of Palghar district.
It is expected that the findings will assist governments 
in making decisions about disaster mitigation and be beneficial to structural 
engineers in designing structures that are resistant to earthquakes.
\section{Seismo-tectonic Setup of the Area of Study}
\label{seistec}
\begin{figure}[!h]
\begin{center}
        \rotatebox{0}{\includegraphics[scale=0.55]{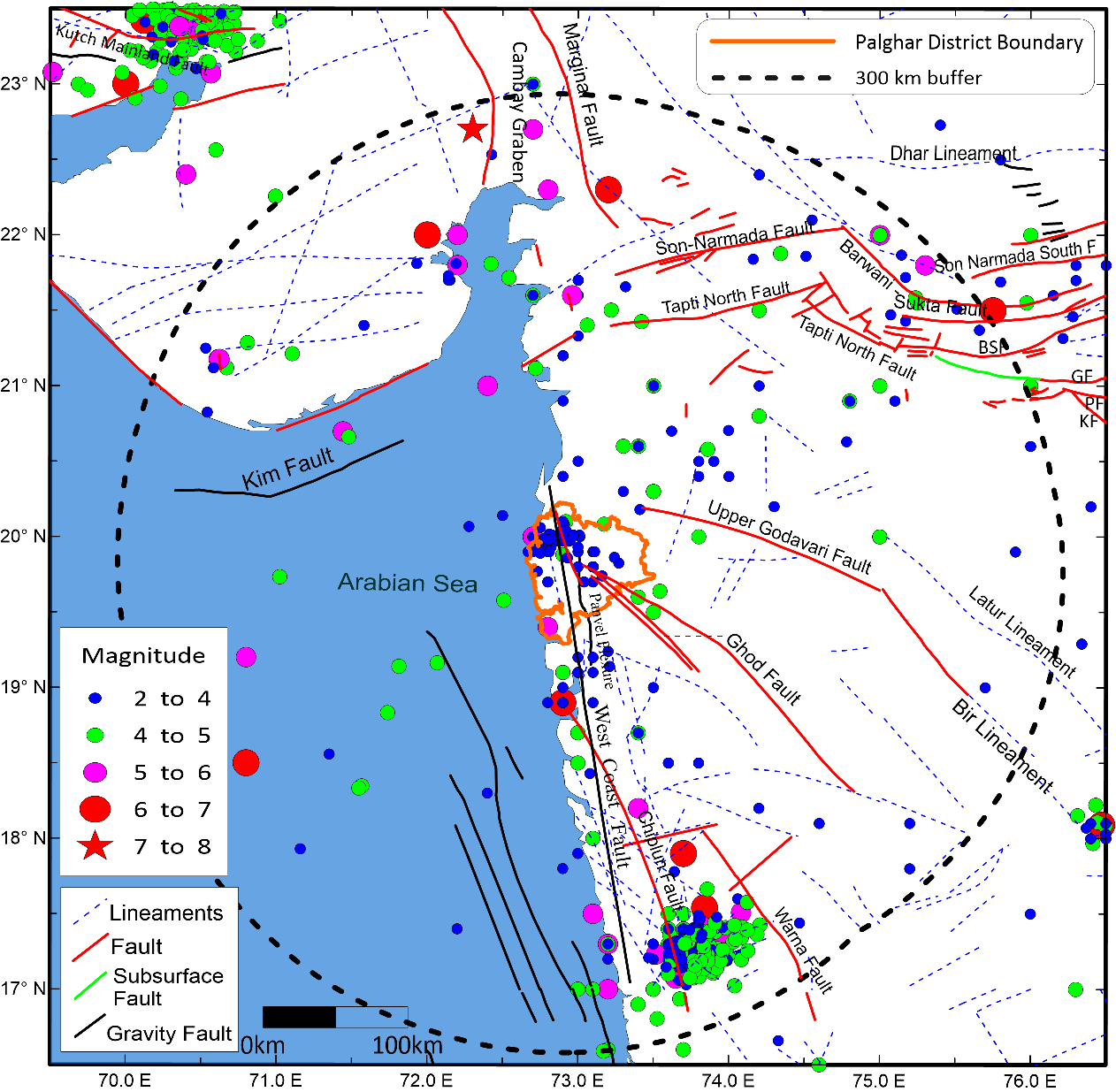}}
        \caption{Map showing the tectonic features with superimposed epicenter of main shocks. 
        A buffer of 300 km (the minumum area considered to estimate the seismicity) from
        the geographical boundary (approximate) of Palghar district is shown
	by dashed black line (GF - Gavilgarh fault, PF - Purna fault, KF - Kadam fault
	and BSF - Barwani-Sukta fault)}
\label{tecmap}       
\end{center}
\end{figure}

Seismic hazard assessment of Palghar district requires evaluating regional seismicity 
within a buffer of 300 km from its geographical boundary. The study 
area is defined as being between $16.5\degree$ N -
$23.5\degree$ N in latitude and $69.5\degree$ E - $76.5\degree$ E in longitude. 
Figure \ref{tecmap} displays the main seismo-tectonic 
features in the study area based on the Seismotectonic Atlas of India and Its 
Environs (SEISAT), which was released by Geological Survey India (GSI) \citep{dasg1}.
The structural trends in the region is comprised of northern
portion of NW trending Western Ghats, Deccan Plateau and
Son Narmada and Tapti Rift Zones. The NW trending Western
Ghats is one of the major uplifted plateaus in the Indian
shield. The Ghats span a wide variety of geological formations with varying structural 
and physical properties. One of the main tectonic features in the 
Ghats is the West-Coast Fault, which is 
naturally the divide between the Indian plate and Gondwanaland. The East Marginal 
Fault of the Cambay Graben may be continuous with this feature. Relative to the 
West Coast Fault, the Chiplun Fault is thought to represent its secondary 
manifestation. One additional significant tectonic feature in this zone is the 
monoclonal Panvel Flexure, which has a western limb that dips towards the west. 
This flexure trends parallel to the West Coast Fault \citep{dasg1} in a NNW direction 
along its axial track. Major faults and lineaments in the Deccan Plateau area trend NW-SE, 
which is in line with the tectonic grains of the nearby Precambrian basement rocks. It 
is discovered that some of the faults and lineaments that cross the trap rocks 
continue into the older rocks that surround them. The Neo Tectonic Fault (Ghod), 
the Upper Godavari Faults (UGF), and a few other faults with basement and cover 
are the main geofractures in this zone. The presence of transverse faults are the 
indication of the abrupt changes in Deccan 
Trap thickness and base level depths. It has been identified that the Son Narmada 
South Fault, which is trending ENE-WSW, is episodically active.
The Narmada North Fault (NNF) and Narmada South Fault (NSF) are the constituents of 
the boundary fault systems that control the Narmada Rift Basin in the area.
These faults extends upto mantle and are reactivated many times in the past. 
The Barwani-Sukta Fault and the Tapti North Fault (TNF) have a parallel 
curved E-W to ENE-WSW appearance. TNF continues along the southern portion of the 
Satpura axial zone, branching out to the east into the Gavilgarh Fault, another ENE-WSW 
facing fault that stretches northeast of Akot and edges out the northern tectonized 
margin of the Purna alluvium.
Some parts of the Narmada South and the TNF exhibit neotectonism. The Son-Narmada Fault
and TNF have ENE-WSW trend, and extends beyondthe Cambay Graben to enter the 
Saurashtra Peninsula towards west. TNF characterizes the southern border of the Satpura range
and northern border of Tapti alluvium. The Kaddam Fault,
gets terminated by the Purna Fault, the southernmost fault of Son-Narmada-Tapti (SONATA)
lineament Zone \citep{dasg1}. To correlate historical seismicity with the existing tectonic 
features, the epicenters of the main shocks are overlaid on the tectonic features
in Figure \ref{tecmap}.
\section{Techniques and Computational Structure}
\label{mcf}
\subsection{Earthquake Catalogue}
The preparation of an earthquake catalogue is the initial step in the assessment of 
seismic hazards. The term "earthquake catalogue" refers to an earthquake database that 
includes information about each individual earthquake event, with its time of 
occurrence (year, month, day, hour, and minute), location  
(longitude, latitude, and depth) 
and magnitude ($M_L$ - local magnitude, $M_S$ - surface 
wave magnitude, $m_b$ - body wave magnitude and $M_W$ - moment magnitude).
For the present analysis, the earthquake catalogue for the instrumental period has been 
prepared using the data available from National Earthquake Information Centre (NEIC), 
United States Geological Survey (USGS), the reviewed International Seismological Centre (ISC) 
bulletin, UK and National Center for Seismology (NSC), India. For pre-instrumental and 
early instrumental period, the data has been obtained from various published sources
\citep{jai1,old2,tond1,rama1}.
There are $714$ events in the catalogue compiled for this study, which begins 
in 1594 A.D. The assembled list includes magnitude in four different scales: 
$M_L$, $M_S$, $m_b$ and $M_W$. Since most GMPEs are developed in terms of $M_W$ to avoid saturation 
effects, conversion of various magnitude scales into one type of magnitude ($M_W$) using 
appropriate conversion relations (called homogenization) is necessary for seismic hazard analysis.
The homogenization has been
carried out using global empirical relations \citep{fliz1,scor1,sipk1}.
The primary shocks and triggered events (foreshocks and aftershocks) are included in the 
seismic events listed in the catalogue. The main shocks have a Poissonian distribution and 
are statistically independent. The triggered events normally depend 
on main shocks and have a tendency to cluster in space and time around the mainshock. 
Assessment of seismic hazards is carried out by assuming that earthquake distribution 
is Poissonian in nature. Consequently, a procedure known as declustering is required 
to be performed to recognize and eradicate the aftershocks and the foreshocks from the catalogue.
The declustering method by Uhrhammer \cite{uhr1} is preferred for Peninsular India and the same 
is adopted here.
There are $635$ main shocks in the declustered earthquake catalogue in $M_W$ units; therefore, 
the current declustered method removes approximately $11$$\%$ events.
\subsection{Demarcation of Seismic Source Zones (SSZ)}
\begin{figure}[!h]
\begin{center}
        \rotatebox{0}{\includegraphics[scale=0.5]{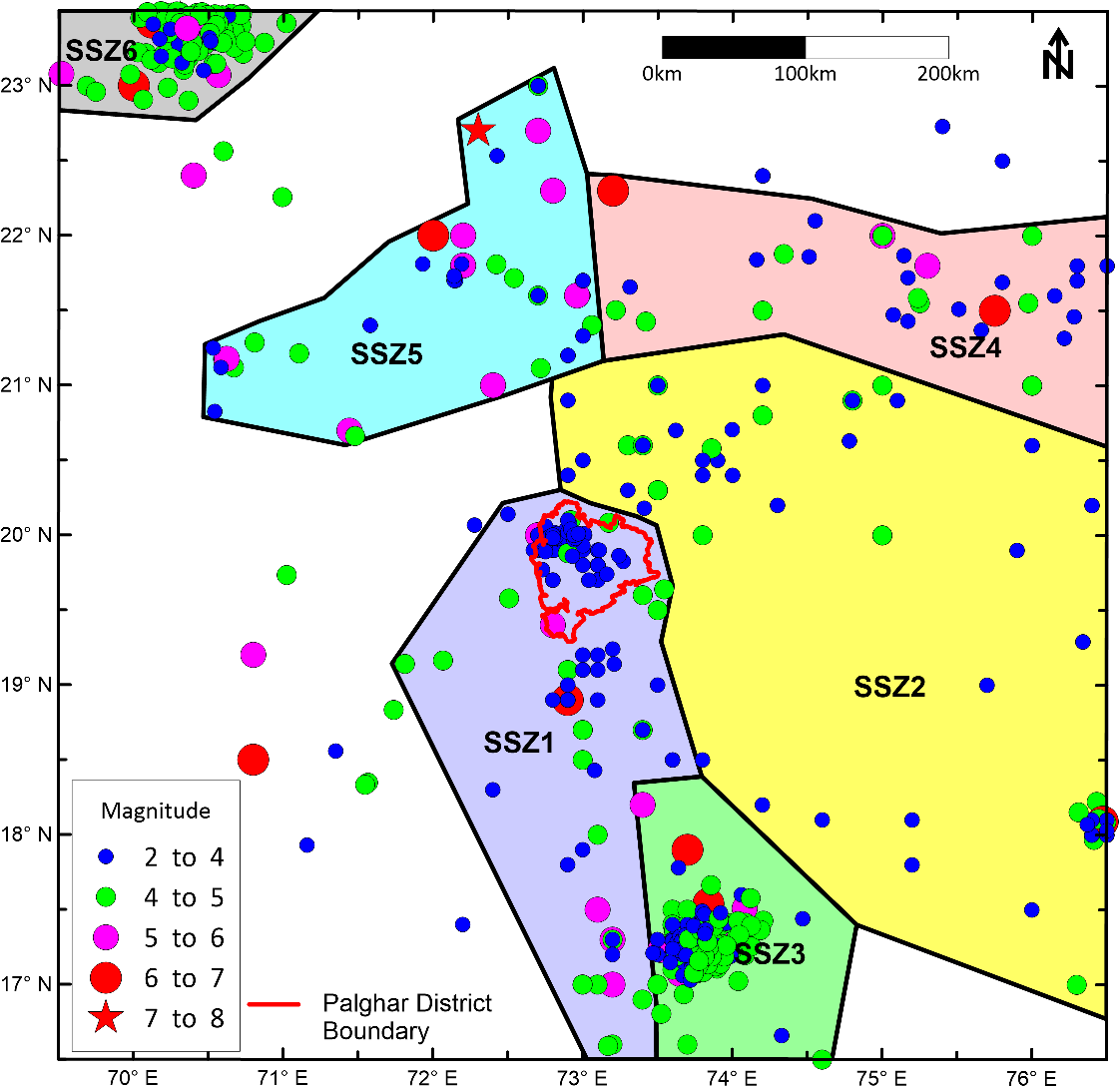}}
        \caption{Seismogenic source zones within the study area
        along with the epicentres of different magnitude ranges}
\label{SSZ}       
\end{center}
\end{figure}

When assessing seismic hazards, one of the most crucial stages is identifying 
seismogenic sources. An area of scattered seismicity with a discernibly diverse 
seismogenic capacity in terms of both the maximum magnitude and the frequency of 
earthquakes in various magnitude ranges is referred to as a SSZ. Six broad SSZs have 
been identified, as illustrated in Figure \ref{SSZ}, taking into account the spatial 
distribution and the correlation between the tectonic features in the study 
area and past seismic activities.
The area covering west coast is designated as
SSZ1 which includes the Palghar district. The region covering
the Upper Godavari fault, Tapti North fault and part of Deccan
Trap is considered as SSZ2 which includes the epicenter
of $M_W$ 6.2 magnitude Killari earthquake of September 29, 1993.
The Koyna region is taken as SSZ3 which includes $M_W$ 6.4 magnitude
Koyna earthquake of December 10, 1967.
SSZ4 represents the SONATA rift zone, SSZ5 falls under the
Sourashtra region and SSZ6 falls under the Kutch region.
\subsection{Completeness of the Catalogue}
\begin{figure}[!h]
\begin{center}
        \rotatebox{0}{\includegraphics[scale=0.6]{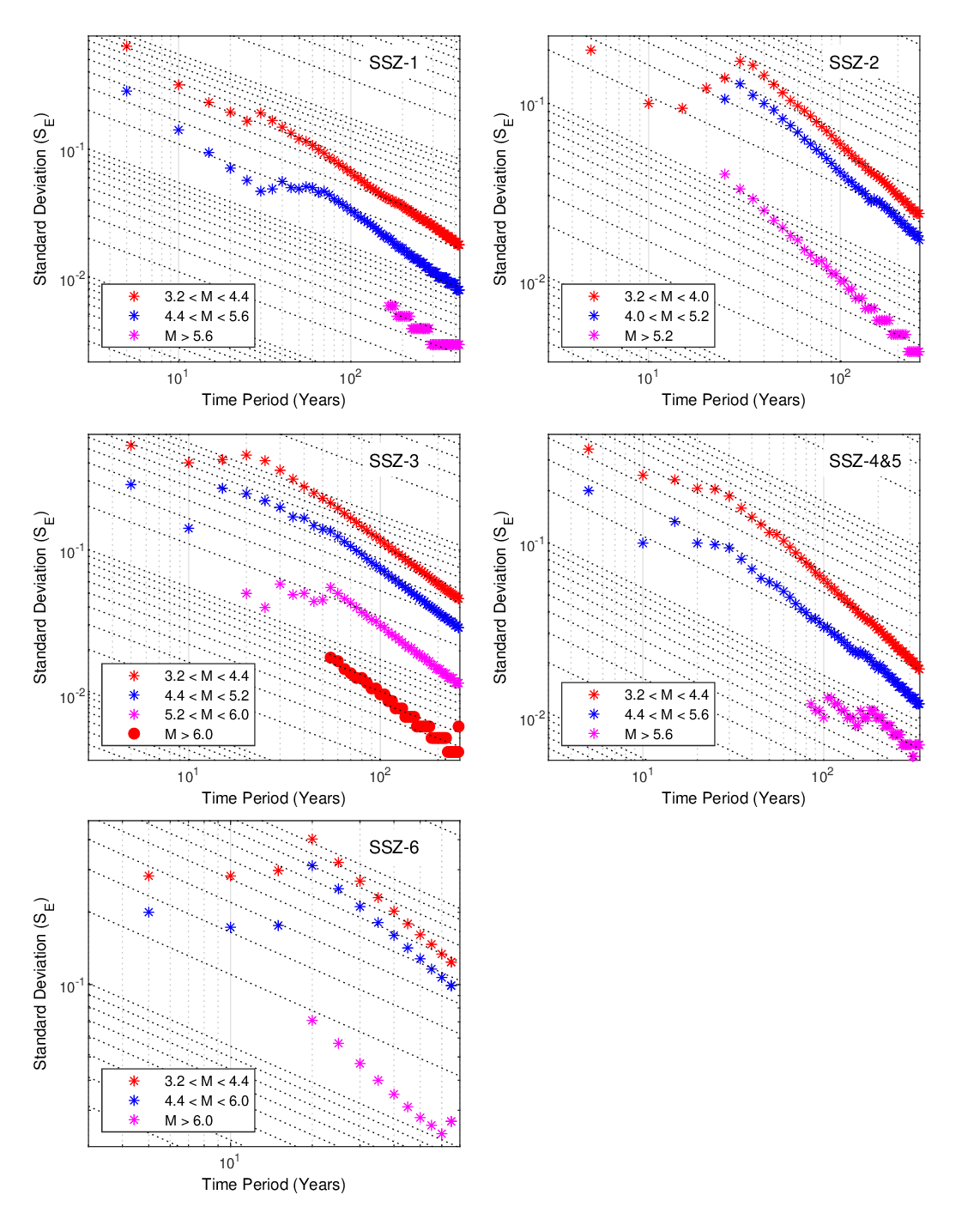}}
        \caption{Plot showing the catalogue completeness for all the SSZs}
\label{stepp}
\end{center}
\end{figure}

It is understood that old, low-magnitude earthquakes with insufficient 
instrumentation often have incomplete data in the catalog. The occurrence 
rates of these events can be assessed using recent data, specifically 
spanning the last $25$ years, because of their short return periods.
But the data for a longer period must be taken 
into account in order to have a reliable estimate for the occurrence 
rates of high-magnitude earthquakes.
The mean rates of earthquake occurrence may be underestimated if 
the data is not corrected for incompleteness. Finding the time period 
of complete data for a pre-set magnitude range will allow 
for the correction effectively. The entire set of data can then be 
used to calculate reliable mean earthquake occurrence rates for the 
specified magnitude ranges.
The statistical method that Stepp \cite{step} proposed to compute the 
completeness period is used in the current analysis. 
According to Stepp, earthquakes will obey a 
Poissonian distribution with an unchanged occurrence rates if all of the 
events are reported in a catalogue.
For a time period of T years, if $E(M)$ is the annual average number of events 
for a particular magnitude interval centered about $M$, the standard 
deviation $S_E$ of $E(M)$ is given by
\begin{equation}
        S_E=\sqrt{E(M)/T}
        \label{catcom}
\end{equation}
The stationarity of $E(M)$ ensures that $S_E$ acts like $1/\sqrt{T}$. A significant 
departure of the $S_E$ values from the linearity of the $1/\sqrt{T}$ slope yields 
the period of completeness from the plot of $S_E$ versus $1/\sqrt{T}$, also 
referred to as the "completeness plot". As the magnitude range increases, 
the period of completeness gets progressively longer.
The completeness plots for all the SSZs are shown in
Figure \ref{stepp}.
\subsection{Determination of $M_{\tt max}$}
The highest seismic event that is typical of a region under seismotectonic 
and geological conditions is known as the maximum earthquake ($M_{\tt max}$). 
There are several techniques for figuring out $M_{\tt max}$ values 
\citep{well1,kijk1,kijk2} and 
each technique always carries some subjectivity \citep{boll1}. For instance, the Wells 
and Coppersmith \cite{well1} method requires the fault rupture length of upcoming 
earthquakes to be specified, but there is insufficient scientific evidence 
to make this determination with any degree of confidence.
The Kijko-Sellevoll-Bayesian (KSB) method, a maximum likelihood method 
for maximum earthquake estimation, has been used in this study due to 
its broad acceptance and strong mathematical foundation. The Bayesian 
version of the Kijko-Sellevoll (KS) estimator of $M_{\tt max}$ \citep{kijk3} can be 
constructed 
using knowledge of the Gutenberg-Richter (GR) probability distribution 
function (PDF), and the KSB estimator of $M_{\tt max}$ can be computed through an 
iterative process. For every SSZ, a maximum earthquake estimation has been made.
\subsection{Determination of Seismicity Parameters}
\begin{figure}[!h]
\begin{center}
        \rotatebox{0}{\includegraphics[scale=0.6]{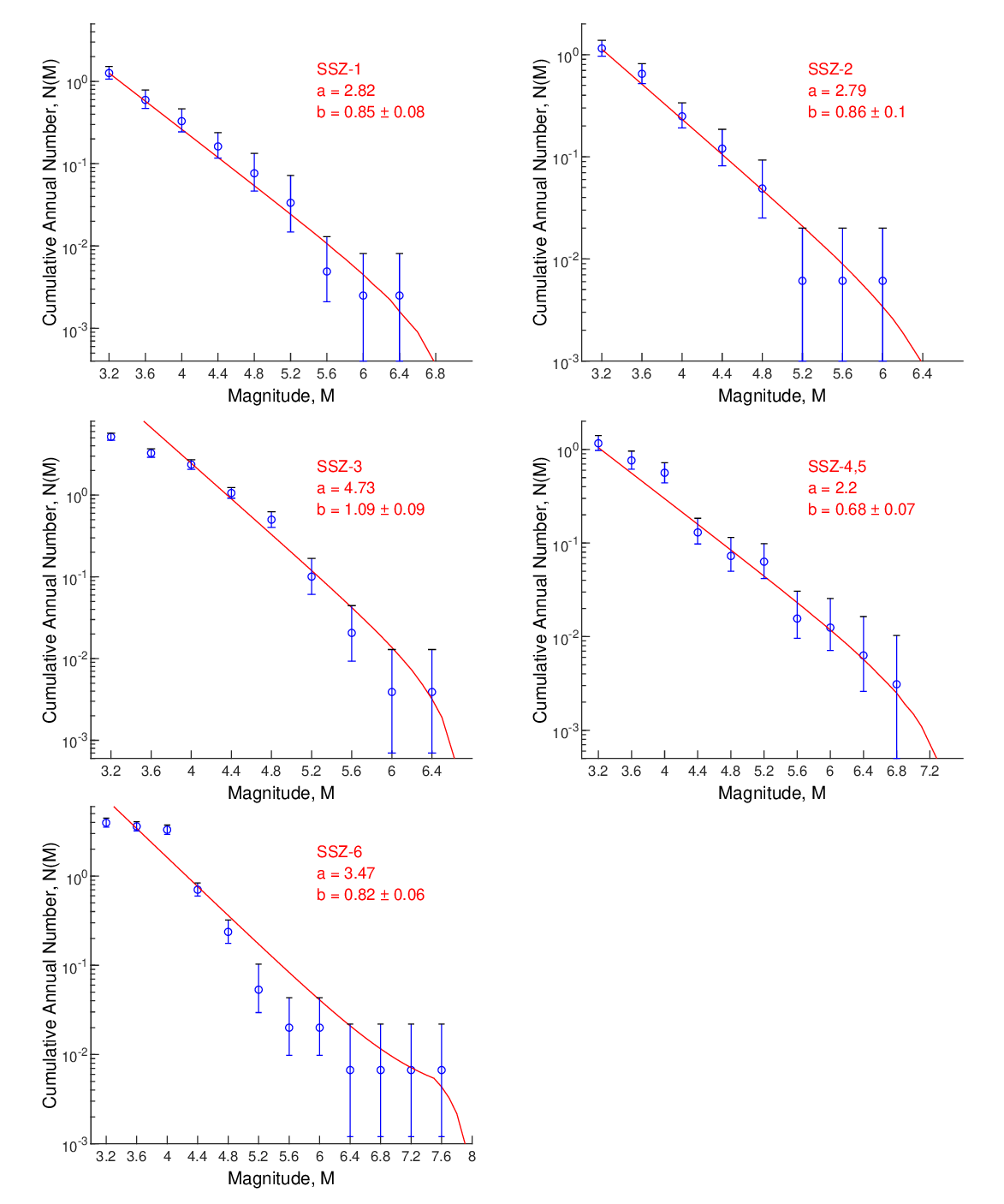}}
\caption{Plot showing the magnitude-frequency curves for all the SSZs.}
\label{recur}
\end{center}
\end{figure}

When estimating seismic hazards, the assessment of seismicity parameters is 
thought to be the most crucial step. It is widely acknowledged that the frequency 
of earthquakes obeys an exponential distribution worldwide, indicated by 
Gutenberg and Richter \cite{guten1} as
\begin{equation}
        \log N(M)=a-bM
        \label{gr}
\end{equation}
where $N(M)$ is the cumulative number of events with magnitude equal to or 
greater than M. The 
mean annual number of earthquakes with a magnitude of zero or higher is $10^a$, 
and the relative likelihood of large and small earthquakes is 
represented by $b$ (or the $b$ value). Regression analysis is typically 
used to estimate the $a$ and the $b$ values from data that is available for a 
SSZ of interest.
The standard Gutenberg-Richter (GR) relation holds good from $-\infty$ to 
$+\infty$ in magnitude range. 
The impact of minor earthquakes are negligible
for engineering purposes, and is generally ignored as it does not
cause any significant damage \citep{kra1}. As a result, 
it is required to set a lower bound, or to take into account a lower 
threshold ($M_{\tt min}$) on the magnitude.
All SSZs, however, are 
always associated with a maximum magnitude ($M_{\tt max}$) or upper bound magnitude. 
The GR law then takes the form of a truncated exponential distribution 
by imposing a lower and an upper bound magnitude, 
in the magnitude range of $M_{\tt min}$ to $M_{\tt max}$ 
and is represented by \citep{corn1}
\begin{equation}
        N(M)=N(M_{\tt min})\frac{e^{-\beta(M-M_{\tt min})}-e^{-\beta(M_{\tt max}-M_{\tt min})}}
        {1-e^{-\beta(M_{\tt max}-M_{\tt min})}}, ~M_{\tt min} < M < M_{\tt max}
        \label{modgr}
\end{equation}
A suitable $M_{\tt min}$ is required for the computation of hazard \citep{bomm1} even 
though the choice of $M_{\tt min}$ in Eqn.~\ref{modgr} is not crucial. $M_{\tt min}$ for this study 
is considered 
as $3.2$ and $\beta=b\ln 10$. Using Weichert's \cite{weic1} maximum likelihood method, the GR 
relationship for a SSZ is fitted by first determining the periods of completeness 
for various magnitude intervals. Due to the extremely low number of events in SSZs $4$ and $5$, 
they are combined to provide a good fit for the truncated GR law.
The magnitude-frequency dependence of SSZ $6$ is not well represented by 
exponential distribution of the truncated 
GR law of Eqn. ~\ref{modgr}. As a result, for this SSZ, a more intricate recurrence 
law called the characteristic earthquake recurrence law, invoked by Youngs and 
Coppersmith \cite{youn2}, is used. Higher rates of exceedance at magnitudes close to the 
characteristic earthquake magnitude and lower rates at lower magnitudes are 
predicted by the characteristic earthquake model. Table \ref{seispar} lists the seismicity 
parameters for each polygonal SSZ.  
Figure \ref{recur} displays the magnitude-frequency distribution 
curves for each SSZ.
\begin{table}[h]
\caption{Computed values of seismicity parameters for all SSZs}
        {\begin{tabular}{cccccc}
\hline\noalign{\smallskip}
 & SSZ & $b$ & $a$ & $M_{\tt max}$ & $M_{\tt max}^{\tt obs}$ \\
\noalign{\smallskip}\hline\noalign{\smallskip}
 &1  & 0.85 $\pm$ 0.08  & 2.82  & 7.04 $\pm$ 0.58 & 6.51  \\
 &2  & 0.86 $\pm$ 0.10  & 2.79  & 6.65 $\pm$ 0.51 & 6.20 \\
 &3  & 1.09 $\pm$ 0.09  & 4.73 & 6.59 $\pm$ 0.31 & 6.40  \\
 &4+5  & 0.68 $\pm$ 0.07  & 2.20 & 7.44 $\pm$ 0.50 & 7.01  \\
 &6  & 0.82 $\pm$ 0.06 & 3.47 & 8.96 $\pm$ 1.29 & 7.70 \\
\noalign{\smallskip}\hline
\end{tabular}
        \label{seispar}}
\end{table}
\subsection{Determination of Smooth-gridded Seismicity}
A physically plausible representation is not offered by the conventional 
uniform spatial distribution model of seismicity, also known as uniformly 
smoothed seismicity. Therefore, the activity rates in each SSZ are 
determined using the smooth-gridded seismicity model \citep{fran1}. 
The annual activity rates in the smooth-gridded seismicity are spatially 
varying, but the $M_{\tt max}$ and the $b$-value stay constant within the source zones. 
According to this, there is no correlation between the $b$-value and the 
activity rate, and the probability of earthquakes within a zone is not 
distributed uniformly \citep{nath2}. Conversely, the uniform areal seismicity hypothesis 
states that there is an equal probability of earthquakes occurring at 
every point in the zone.

\begin{figure}[!h]
\begin{center}
        \rotatebox{0}{\includegraphics[scale=0.6]{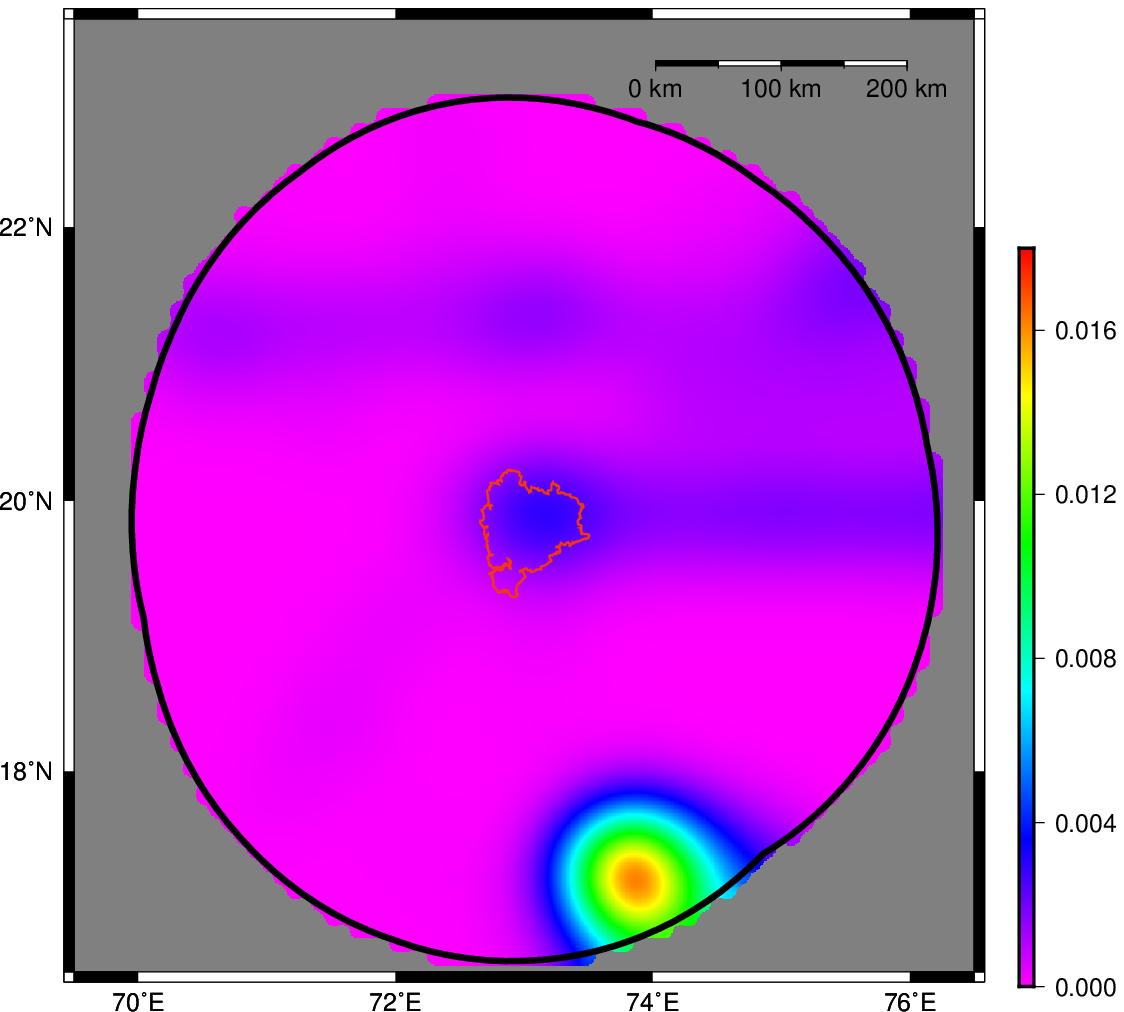}}
\caption{Map showing the smoothened annual activity rate for 
	threshold magnitude $3.2$ $M_W$}
\label{sgs}
\end{center}
\end{figure}

The discrete earthquake distributions can be modeled into spatially continuous 
probability distributions using the smooth-gridded seismicity model. In the 
current study, the method provided by Frankel \cite{fran1} is used for this purpose. 
Several researchers have previously used this 
technique \citep{fran2,stir1,lap1,jai2,nath4,nath2}. The 
following is the smoothed function:
\begin{equation}
        N_i(m_r)=\frac{\sum\limits_{j} n_j(m_r)e^{-(d_{ij}/c})^2}
        {\sum\limits_{j} e^{-(d_{ij}/c})^2}
\label{sgseq}
\end{equation}
where $c$ indicates the correlation distance, which is related to the uncertainty 
in the epicentral location and is assumed to be $50$ km in this 
analysis, $d_{ij}$ is the distance between $i^{th}$ and $j^{th}$ cells, and $n_j(m_r)$ is 
the number of events with magnitude $\geq m_r$. The sum is computed in 
cells $j$ within a distance of $3c$ of cell $i$. For threshold magnitude $3.2$ $M_W$, 
the annual activity rate $\lambda_{m_r}$ is calculated as $N_i(m_r)/T$, where $T$ is 
the (sub)catalogue period. The period of years $1990 - 2021$ are covered 
by the sub-catalogue for the threshold magnitude of $3.2$ $M_W$. Figure \ref{sgs} shows 
the smoothened annual activity rate for threshold magnitude $3.2$ $M_W$. It is 
possible to identify the areas of expected asperities using the 
smooth-gridded seismicity analysis \citep{nath4}.
\subsection{Data-driven methods for suitability check of GMPEs}
Selecting and prioritizing GMPEs based on recorded strong motion 
data is crucial for effectively applying the logic tree method in the 
PSHA to integrate epistemic uncertainties. A comprehensive and in-depth 
quantitative analysis has been conducted to determine the most appropriate GMPEs in 
the study area. To this end, sixteen GMPEs 
were selected and their performance was compared to the observation (the 
strong motion data that was recorded). There are multiple approaches 
to assess the goodness-of-fit metrics \citep{sche1,sche2,kale1,kowa1}.
The Likelihood (LH) method is one of them which is a 
transparent one, based on the concept of likelihood \citep{edwa1}. 
However, the $LH$ method is subject to certain subjectivities, 
even though it is statistically significant. Its effectiveness depends 
on the sample size, or the total amount of strong motion data that have 
been recorded. Subjectivities are found in the definition of ranks and 
in the threshold values of the goodness-of-fit measures. 
Scherbaum {\it et al.} \cite{sche2} suggested the log-likelihood 
(LLH) method, a different 
goodness-of-fit technique that is independent of \textit{ad hoc} assumptions, in 
order to get around these drawbacks.

Information theory ideas serve as the foundation for the $LLH$ approach. Measuring 
the Kullback-Leibler (KL) distance between two models - a model that represents 
reality (in this case, the ground motion data) and a candidate ground motion 
model (in this case, a GMPE) is the main goal of the $LLH$ method. When two 
probability density functions, $f(x)$ and $g(x)$, describe two models, 
$f$ and $g$, the KL distance between them is expressed as \citep{sche2}
\begin{equation}
        D(f,g) = E_{f}[\log_2 (f)] - E_{f}[\log_2 (g)]
\label{KLd}
\end{equation}
where $E_{f}$ statistical expectation of the log-likelihood of the model function 
with respect to $f$. $D(f,g)$ denotes the amount of information 
lost if model $f$ is replaced by model $g$ and is understood as the 
relative entropy between $f$ and $g$ \citep{cove1}. A natural distance measure is 
indicated by a negative value of the  
statistical expectation \citep{sche2} of the log-likelyhood function. 
The relative KL distance is the only measure 
that matters when comparing models because the statistical expectation 
of the log-likelihood of $f$ with respect to $f$ cancels out as a constant 
term \citep{sche2}. The log-likelihood score can therefore be defined as follows in 
order to define a ranking criterion:
\begin{equation}
        LLH = -\frac{1}{N} \sum_{n=i}^{N} \log_2(g(x_i))
\label{llheq}
\end{equation}
$N$ is the total number of observations in this case. This LLH score is called 
average sample log-likelihood. More explicitly, the 
LLH score in Eqn. \ref{llheq} is expressed as
\begin{equation}
        LLH = -\frac{1}{N} \sum_{n=i}^{N} \log_2 \left[\frac{1}{\sigma_i
        \sqrt{2\pi}}\exp\left\{-\frac{1}{2}\left(\frac{y_i-\mu_i}
        {\sigma_i}\right)^2\right\}\right]
\label{llheq1}
\end{equation}
Here, the observed ground motion value is denoted by $y_i$, and the mean and 
standard deviation of the GMPE under consideration are represented 
by $\mu_i$ and $\sigma_i$. A GMPE that performs better for a target region is 
indicated by a lower LLH score. With the LLH approach, a GMPE's suitability 
can be evaluated using just one metric - the LLH score, which essentially 
represents the average information loss that occurs when a candidate GMPE 
replaces the model that represents reality. On the other hand, the LH method 
requires the combination of four measures in order to determine the 
ranking criteria. Thus, the LLH approach is considered to 
be superior then LH approach.
\subsection{Ground motion data set}
The region in which the study area is located has limited strong motion 
data available.
A set of $66$ three-component strong-motion accelerograms
($132$ horizontal accelerograms of longitudinal and transverse
components) avilable from $53$ earthquakes have been considered to perform 
the suitabilty test of GMPEs. Out of these $53$ earthquakes,
$51$ haven been reported in Koyna region, Maharashtra and the other two
earthquakes took place at Bhuj, Gujarat and at Osmanabad, Maharashtra.
The Bhuj earthquake (January 26, 2001) was recorded at Ahmedabad Station,
Gujarat and all other earthquakes were recorded at Koyna Dam
(Shear Zone Gallery, 1A Gallery, 1B Gallery, Koyna Dam Downstream,
Koyna Dam Top, Kirnos Observatory, Middle Gallery and Koyna Dam Bottom).
The earthquakes considered are listed in Table \ref{smeqs}.
The strong-motion data, used for performing
the efficacy test of GMPEs, have been taken from
the records available at Central Water and Power Research Station (CWPRS),
Pune and few have been downloaded from Center for Engineering Strong Motion Data.

{\fontsize{8}{9}\selectfont
\begin{longtable}{lccclcc}
\caption{List of earthquakes used for performing
        the suitability analysis} \\
\hline\noalign{\smallskip}
        Serial & Date & Lat & Lon & $M_W$ & $H$ & $R_{epi}$ \\
   No.  & (dd/mm/yy) & ($\degree N$) & ($\degree E$) & & (km) & (km) \\
\noalign{\smallskip}\hline\noalign{\smallskip}
  1.  & 12/09/1967  & 17.43  & 73.72  & 3.9  &  3.0  &   4.9  \\
  2.  & 13/09/1967  & 17.40  & 73.70  & 4.5  &  5.0  &   5.5  \\
  3.  & 13/09/1967  & 17.36  & 73.76  & 3.2  &  7.0  &   4.3  \\
  4.  & 16/11/1967  & 17.45  & 73.85  & 3.5  &  8.0  &  12.0  \\
  5.  & 10/12/1967  & 17.51  & 73.73  & 6.5  & 12.0  &  12.0  \\
  6.  & 11/12/1967  & 17.30  & 73.89  & 3.8  &  8.0  &  18.5  \\
  7.  & 12/12/1967  & 17.40  & 73.76  & 3.6  &  3.0  &   1.1  \\
  8.  & 12/12/1967  & 17.28  & 73.69  & 4.7  & 13.0  &  14.1  \\
  9.  & 13/12/1967  & 17.30  & 73.78  & 4.6  & 15.0  &  11.7  \\
 10.  & 13/12/1967  & 17.49  & 73.78  & 3.8  & 23.0  &   9.8  \\
 11.  & 14/12/1967  & 17.31  & 73.78  & 4.1  & 12.5  &  10.7  \\
 12.  & 14/12/1967  & 17.37  & 73.75  & 4.1  &  5.0  &   3.0  \\
 13.  & 17/12/1967  & 17.31  & 73.75  & 3.7  &  3.0  &  10.0  \\
 14.  & 24/12/1967  & 17.35  & 73.71  & 5.0  & 20.0  &   6.9  \\
 15.  & 24/12/1967  & 17.35  & 73.71  & 5.0  & 20.0  &   6.9  \\
 16.  & 12/01/1968  & 17.39  & 73.75  & 4.1  &  4.0  &   1.5  \\
 17.  & 12/01/1968  & 17.39  & 73.75  & 4.1  &  4.0  &   1.5  \\
 18.  & 14/02/1968  & 17.33  & 73.70  & 3.6  & 16.0  &   8.9  \\
 19.  & 04/03/1968  & 17.36  & 73.77  & 4.2  & 10.0  &   4.0  \\
 20.  & 04/03/1968  & 17.36  & 73.77  & 4.2  & 10.0  &   4.0  \\
 21.  & 29/10/1968  & 17.35  & 73.72  & 5.2  &  6.5  &   6.0  \\
 22.  & 29/10/1968  & 17.35  & 73.72  & 5.2  &  6.5  &   6.0  \\
 23.  & 29/10/1968  & 17.35  & 73.72  & 5.2  &  6.5  &   6.0  \\
 24.  & 27/06/1969  & 17.40  & 73.74  & 3.9  &  3.0  &   1.8  \\
 25.  & 27/06/1969  & 17.40  & 73.74  & 4.7  &  3.0  &   1.1  \\
 26.  & 01/01/1970  & 17.33  & 73.71  & 4.3  & 11.0  &   9.0  \\
 27.  & 27/05/1970  & 17.48  & 73.81  & 4.4  &  3.0  &   9.0  \\
 28.  & 17/06/1970  & 17.32  & 73.31  & 3.6  &  1.0  &  49.1  \\
 29.  & 17/06/1970  & 17.32  & 73.31  & 3.6  &  1.0  &  49.1  \\
 30.  & 26/09/1970  & 17.37  & 73.65  & 4.4  & 13.0  &  11.0  \\
 31.  & 14/02/1971  & 17.36  & 73.83  & 4.2  &  3.0  &   9.0  \\
 32.  & 17/02/1974  & 17.25  & 73.76  & 4.7  & 19.0  &  16.0  \\
 33.  & 29/05/1974  & 17.49  & 73.78  & 3.5  & 11.0  &  11.0  \\
 34.  & 29/07/1974  & 17.32  & 73.75  & 4.3  & 24.0  &   8.0  \\
 35.  & 02/09/1975  & 17.36  & 73.69  & 4.0  &  7.0  &   8.0  \\
 36.  & 14/03/1976  & 17.24  & 73.73  & 3.9  &  5.0  &  18.0  \\
 37.  & 22/04/1976  & 17.36  & 73.68  & 3.8  & 13.0  &   9.0  \\
 38.  & 12/12/1976  & 17.38  & 73.73  & 3.9  & 13.0  &   3.0  \\
 39.  & 19/09/1977  & 17.27  & 73.75  & 4.0  & 19.0  &  13.0  \\
 40.  & 02/09/1980  & 17.24  & 73.74  & 4.3  &  6.0  &  17.5  \\
 41.  & 02/09/1980  & 17.24  & 73.74  & 4.3  &  6.0  &  17.5  \\
 42.  & 20/09/1980  & 17.21  & 73.76  & 4.7  &  8.0  &  21.0  \\
 43.  & 20/09/1980  & 17.25  & 73.70  & 4.9  &  8.0  &  17.0  \\
 44.  & 20/09/1980  & 17.25  & 73.70  & 4.9  &  8.0  &  17.0  \\
 45.  & 26/10/1980  & 17.25  & 73.74  & 3.7  & 13.0  &  16.0  \\
 46.  & 25/01/1981  & 17.30  & 73.73  & 3.7  &  7.0  &  11.0  \\
 47.  & 25/04/1982  & 17.24  & 73.70  & 4.3  & 13.0  &  18.0  \\
 48.  & 25/04/1982  & 17.24  & 73.70  & 4.3  & 13.0  &  18.0  \\
 49.  & 14/11/1984  & 17.24  & 73.78  & 4.4  &  8.0  &  18.0  \\
 50.  & 14/11/1984  & 17.24  & 73.78  & 4.4  &  8.0  &  18.0  \\
 51.  & 29/10/1989  & 17.32  & 73.77  & 4.0  &  4.0  &   9.0  \\
 52.  & 18/08/1993  & 17.34  & 73.75  & 3.6  &  5.0  &   6.0  \\
 53.  & 28/08/1993  & 17.21  & 73.73  & 4.9  & 12.0  &  21.0  \\
 54.  & 28/08/1993  & 17.20  & 73.78  & 4.9  & 12.0  &  20.0  \\
 55.  & 03/09/1993  & 17.21  & 73.75  & 4.7  & 14.0  &  21.0  \\
 56.  & 29/09/1993  & 18.11  & 76.60  & 6.3  & 15.0  & 321.0  \\
 57.  & 08/12/1993  & 17.17  & 73.72  & 5.1  & 10.0  &  26.0  \\
 58.  & 08/12/1993  & 17.20  & 73.76  & 5.1  &  8.0  &  21.0  \\
 59.  & 01/02/1994  & 17.31  & 73.72  & 5.4  & 12.0  &  10.0  \\
 60.  & 12/03/1995  & 17.25  & 73.73  & 4.7  &  5.0  &  17.0  \\
 61.  & 13/03/1995  & 17.22  & 73.72  & 4.4  &  5.0  &  19.0  \\
 62.  & 25/04/1997  & 17.35  & 73.76  & 4.4  &  3.0  &   6.0  \\
 63.  & 12/03/2000  & 17.20  & 73.72  & 5.2  & 12.0  &  22.0  \\
 64.  & 09/05/2000  & 17.17  & 73.76  & 3.9  &  2.0  &  25.0  \\
 65.  & 10/09/2000  & 17.21  & 73.74  & 3.9  &  5.0  &  21.0  \\
 66.  & 26/01/2001  & 23.42  & 70.23  & 7.7  & 16.0  & 239.0  \\
\noalign{\smallskip}\hline
\label{smeqs} \\
\end{longtable}
}
\subsection{GMPEs used for suitability test}
\begin{figure}[!h]
\begin{center}
        \rotatebox{0}{\includegraphics[scale=0.6]{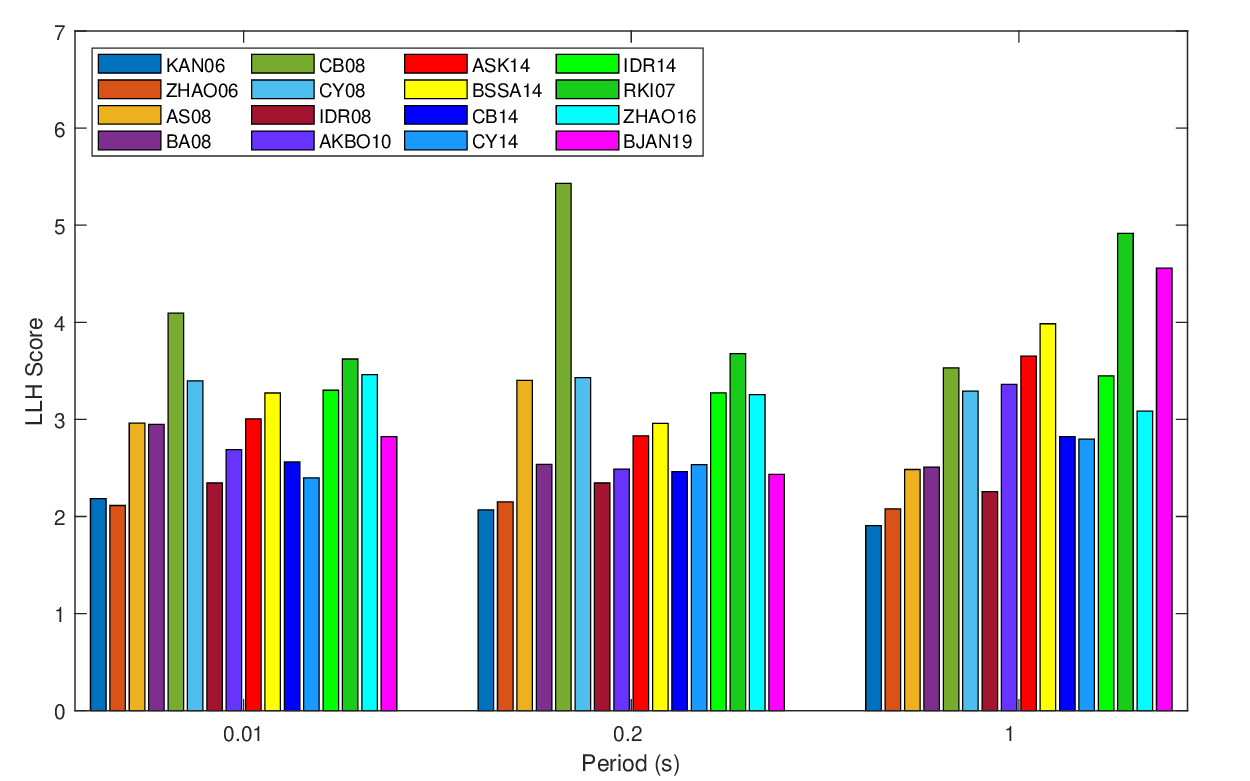}}
\caption{Chart showing the values of LLH score at 0.01s, 0.2s and 1.0s for all the GMPEs
used in the study.}
\label{llhbar}
\end{center}
\end{figure}

Since 2000, there has been a significant increase in seismic networks, leading 
to significant advancements in the development of GMPEs. The report by 
Douglas \cite{doug1} provides a comprehensive and well-documented account of the 
empirical GMPEs that were in existence globally between 1964 and 2021. For the 
purpose of conducting the suitability test and ranking the GMPEs, 16 GMPEs developed 
for crustal earthquakes were primarily selected in accordance with the minimal 
criteria prescribed by Cotton \textit{et al.}\cite{cott1} and Bommer and 
Crowley \cite{bomm1}. The majority of 
the GMPEs are relatively robust, adequately constrained and latest. Twelve of the 
sixteen GMPEs were developed using the global data, and the 
other four, including one for the Himalayan region, were developed for 
other regions. Table \ref{gmpetab} contains a list of all the GMPEs along with their range 
of applicability. 
It is important to note that India's small seismic network and lack of ground 
motion data make it difficult to create reliable GMPEs through empirical means, 
at least not for the study areas. Additionally, it is added 
that, from the observation of Cotton \textit{et al.}\cite{cott1}, a GMPE developed for a region 
that does not tectonically conform to the region under study should be 
excluded is not possibly correct because a GMPE can never be selected or 
excluded based on geographic criteria \citep{byko1,bomm1}.
With a few notable exceptions in active regions \citep{stra1} for high-frequency response 
spectra, multiple studies demonstrate that there is no conclusive evidence of 
regional variations in ground motions in the regions of similar tectonic nature, at least 
from medium to large magnitude earthquakes \citep{staf1,doug2}. Instead, this criterion 
should be understood to mean that the GMPEs for subduction earthquakes should 
not be included in the hazard calculations for crustal earthquakes, and 
vice versa. For instance, GMPEs designed for volcanic areas should not be applied 
to areas that do not experience volcanic activity. As a result, the GMPEs shown 
in Table \ref{gmpetab} have been taken into consideration for the target regions following 
the preliminary analysis.
\begin{table}[!h]
        \caption{GMPEs used for suitability test with their range
        of applicability}
\label{gmpetab}
\begin{adjustbox}{max width=1.0\textwidth,center}
\begin{tabular}{lccccl}
\hline\noalign{\smallskip}
        GMPEs with & Magnitude & Distance & $V_{s30}$ & Periods & Region   \\
        references & ($M_W$) range & range (km) & range (m/s) & (s) & \\
\noalign{\smallskip}\hline\noalign{\smallskip}
        KAN06 (\cite{kann1}) & 5.5 - 8.2 & $<$ 450 & 150 - 1500 & 0.0 - 5.0 & Japan \\
        ZHAO06 (\cite{zhao06}) &       5 - 7.5 &       0 - 300 &       4 site classes  &       0.0 - 5.0       &       Japan \\
        AS08 (\cite{abra2}) & 5 - 8.5  & $<$ 200 & $>$ 180 & 0.0 - 10 & WW \\
        BA08 (\cite{boore1}) & 5 - 8.0 & 0 - 200 & 180 - 1300 & 0.0 - 10 & WW \\
        CB08 (\cite{cam4}) & 4 - 8.5 & 0 - 200 & 150 - 1500 & 0.0 - 10 & WW \\
        CY08 (\cite{chi1}) & 4 - 8.5 & 0 - 200 & 150 -1500 & 0.0 - 10 & WW \\
        IDR08 (\cite{idr08})   &       4.5 - 8 &       0 - 200 & $>$ 450       &       0.0 - 10        &       WW \\
        AKBO10 (\cite{akbo10}) & 5 - 7.6 & $<$ 100 & 3 site classes & 0 - 3.0 & MME \\
        ASK14 (\cite{ask14})   &       3 - 8.5 &       0 - 300 &       180 - 1500      &       0.0 - 10        &       WW \\
        BSSA14 (\cite{bssa14}) & 3.0 - 8.5 (S, R) 3.3 - 7.0 (N) & 0 - 400   & 150 - 1500 &  0.0 - 10 & WW \\
        CB14 (\cite{cb14}) & 3.3 - 8.5 (S)     3.3 - 8.0 (R)    3.3 - 7.0 (N) & 0 - 300 & 150 - 1500 & 0.01 - 10 &  WW \\
        CY14 (\cite{cy14}) & 3.5 - 8.5 (S)    3.5 - 8.0 (R, N) &    0 - 300 & 180 - 1500 &  0.0 - 10 & WW \\
        IDR14 (\cite{idr14})   &       5 - 8.0 &       0 - 150 &       450 - 2000      &       0.0 - 10        &       WW \\
	RKI07 (\cite{raghu1})     &       5.0 - 8.0       &       30 - 300 & 4 site classes & 0.0 - 4.0 & Peninsular India \\
        ZHAO16 (\cite{zhao16c}) &      5 - 7.5 & 0 - 300       &       4 site classes  &       0.0 - 5.0       &       WW \\
        BJAN19 (\cite{bjan19}) &       4 - 9.0 &       10 - 750        &       NA      &       0.0 - 10        &       Himalayas \\
\noalign{\smallskip}\hline\noalign{\smallskip}
        * S = Strike-slip, R = Reverse, N = Normal; & MME = Mediterranean, Middle East and Europe, WW = World Wide  & & & & \\
\noalign{\smallskip}\hline
\end{tabular}
\end{adjustbox}
\end{table}
For ease of understanding and visualization, the bar diagrams of LLH scores at 
three relevant periods for PSHA for each of the GMPEs considered 
are shown in \ref{llhbar}. 
Because of their lowest LLH scores, Fig. \ref{llhbar} makes it clear that the GMPEs 
attributed to KAN06, ZHAO06, and IDR08 are the most appropriate for the 
current study region. Consequently, for the purposes of this study, a combination 
of these three GMPEs with suitable weight factors has been chosen. The weight factors
have been decided as per the prescription given in Scherbaum {\it et al.} \cite{sche2}.
Table \ref{llhsc} 
provides the LLH scores at the relevent periods as well as the weight factors 
assigned to each of these GMPEs.
\begin{table}[h]
\caption{The LLH scores of the selected GMPEs at three periods of importance
	with respective weight factor}
        {\begin{tabular}{ccccc}
\hline\noalign{\smallskip}
		 GMPEs &  &  LLH Score &              & Weight Factor\\
		 selected & 0.01s & 0.20s & 1.0s  & \\
\noalign{\smallskip}\hline\noalign{\smallskip}
ZHAO06 & 2.11  & 2.05  & 2.06 &  0.36   \\
KAN06 &  2.18  & 2.19  & 2.02 &  0.34   \\
IDR08 &  2.35  & 2.24  & 2.20 &  0.30   \\
\noalign{\smallskip}\hline
\end{tabular}
	\label{llhsc}}
\end{table}
\subsection{Hazard Computation}
\label{ct}
Ground motion level is typically used to quantify the seismic hazard at a 
given location. The aggregate hazard at a specific location can be expressed 
by combining the seismic hazard values for each individual SSZ. The PSHA 
methodology takes into account the frequency with which the annual rate of 
ground motion at a given site of interest exceeds a given value for different 
return periods of the hazard. For a given possible earthquake  
magnitude at a source-to-site distance, the 
probability of exceeding a given value of $y^*$ of a ground motion parameter $Y$ is 
calculated, and it is then multiplied by the probability that the earthquake of 
that specific magnitude would occur at that specific location. After that, the 
procedure is repeated for every possible location and magnitude, with the 
probabilities of each being added up.

The total probability theorem \citep{kra1} is used to calculate the probability that a 
ground motion parameter $Y$ will exceed a specific value $y^*$ for a given 
earthquake occurrence.
\begin{equation}
P\left[Y>y^*\right]=P\left[Y>y^*\vert \bf{X}\right]P\left[\bf{X}\right]
=\int P\left[Y>y^*\vert \bf{X}\right] f_x\left(\bf{X}\right)dx
\label{tph}
\end{equation}
where $\bf{X}$ is a vector of random variables that influences $Y$. In order 
to compute the seismic hazard, the quantities in $\bf{X}$ are
limited to magnitude ($M$) and distance ($R$). Assuming that $M$ and $R$
are independent, the probability of exceedance can be written as
\begin{equation}
P\left[Y>y^*\right]=\iint P\left[Y>y^*\vert m,r\right]f_M(m)f_R(r)~dm~dr
\label{poe1}
\end{equation}
where $f_M(m)$ and $f_R(r)$ are the probability density
functions (pdf) and $P\left[Y>y^*\vert m,r\right]$ is obtained from the predictive
relationship for $M$ and $R$ respectively.
While the site is in a region of $N_s$ number of SSZs, and each SSZ has
an average rate of threshold magnitude exceedance (or annual activity
rate) $\lambda_i$, then, the total average exceedance rate for the region is
given by
\begin{equation}
\nu_{y^*}=\sum_{i}^{N_s}\lambda_i \iint P\left[Y>y^*\vert m,r\right]
        f_{Mi}(m)\,f_{Ri}(r)\,dm\,dr
\label{aer1}
\end{equation}
For almost all realistic PSHAs, the integral \ref{aer1} cannot be evaluated 
analytically. This means that numerical integration is necessary. The approach, 
generally employed, is to discretize the possible distance and magnitude 
ranges into $N_M$ and $N_R$ segments, respectively. Now, one can estimate the average 
exceedance rate by
\begin{equation}
        \nu_{y^*} = \sum_{i=1}^{N_s} \sum_{j=1}^{N_M} \sum_{k=1}^{N_R}
        \lambda_i P\left[Y>y^*\vert m_j,r_k\right] f_{Mi}(m_j)\,
        f_{Ri}(r_k)\,\Delta m\,\Delta r
\label{aer2}
\end{equation}
Assuming that each SSZ is able to produce $N_M$ distinct earthquakes with a magnitude 
of $m_j$ at $N_R$ distinct source-to-site distances $r_k$, Eqn. \ref{aer2} can take the form as
\begin{equation}
        \nu_{y^*} = \sum_{i=1}^{N_s} \sum_{j=1}^{N_M} \sum_{k=1}^{N_R}
        \lambda_i P\left[Y>y^*\vert m_j,r_k\right]
        P \left[M=m_j \right] P \left[R=r_k \right]
\label{aer3}
\end{equation}
The return period for the ground
motion parameters $y^*$ can be obtained by taking the reciprocal of $\nu_{y^*}$. 
PSHA intends to generate a seismic hazard curve, which is represented as a 
plot of $\nu_{y^*}$ against $y^*$. The relationship between a ground motion parameter 
and its frequency of exceedance is provided by the seismic hazard curve.
The probability of exceedance of $y^*$ in a
finite time interval (say, for an exposure period $T$) can be estimated by combining 
the seismic hazard curve with the Poisson model and is given by
\begin{equation}
P \left[Y_T > y^* \right]=1 - e^{-\nu_{y^*}T}
\label{pom}
\end{equation}
The PSHA computational method discussed above is applied to every grid 
point (site) at the engineering bedrock level ($V_{S30} \sim
760\, m/s$). Using the framework provided by Kaklamanos \textit{et al.}\cite{kakl1} for 
estimating unknown input parameters, the 
various distance parameters, such as rupture distance ($R_{rup}$), horizontal distance 
to top edge of the rupture ($R_x$), etc., needed for implementing the GMPEs, have 
been determined.
\section{Results and Discussion}
\label{rd}
\begin{figure}[!h]
\begin{center}
\begin{tabular}{cc}
      \resizebox{65mm}{!}{\rotatebox{0}{\includegraphics[scale=0.6]{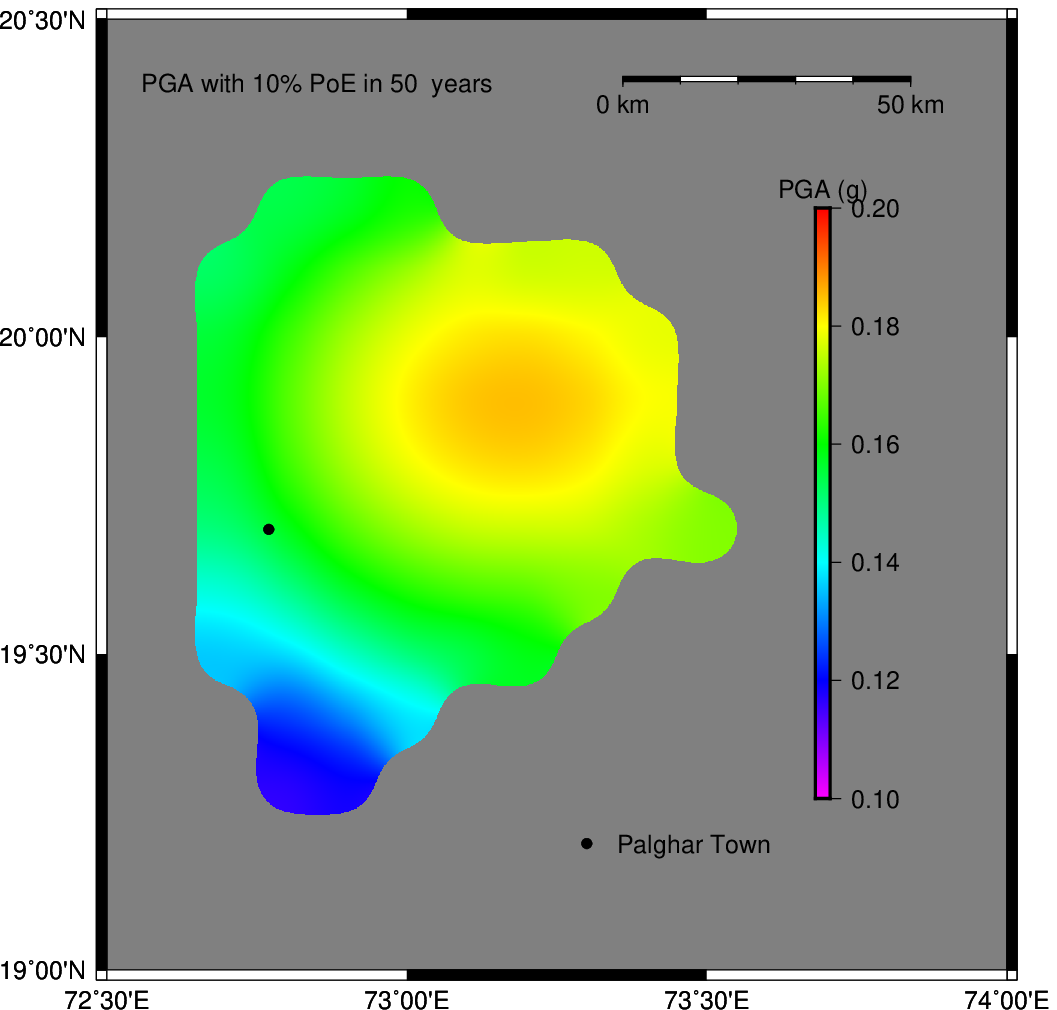}}} &
      \resizebox{65mm}{!}{\rotatebox{0}{\includegraphics[scale=0.6]{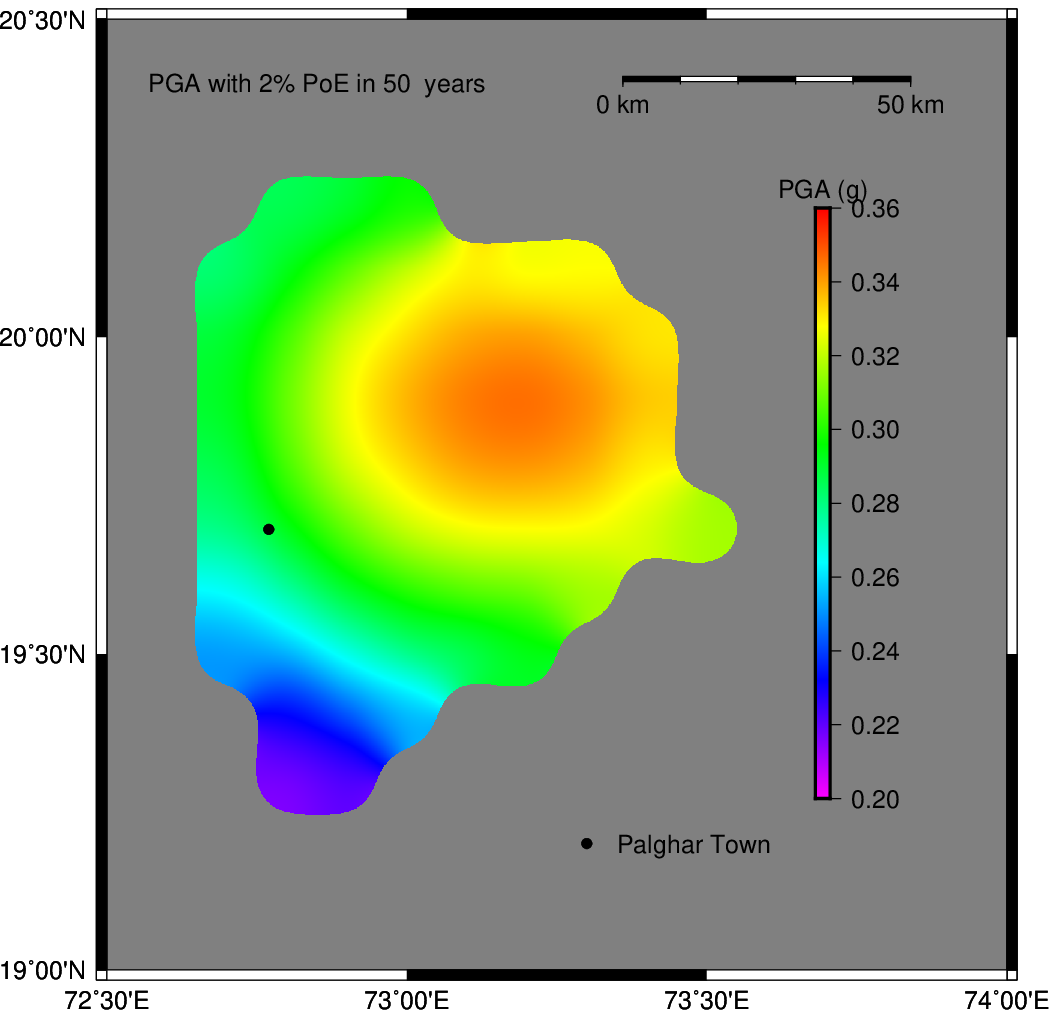}}} \\
      \resizebox{65mm}{!}{\rotatebox{0}{\includegraphics[scale=0.6]{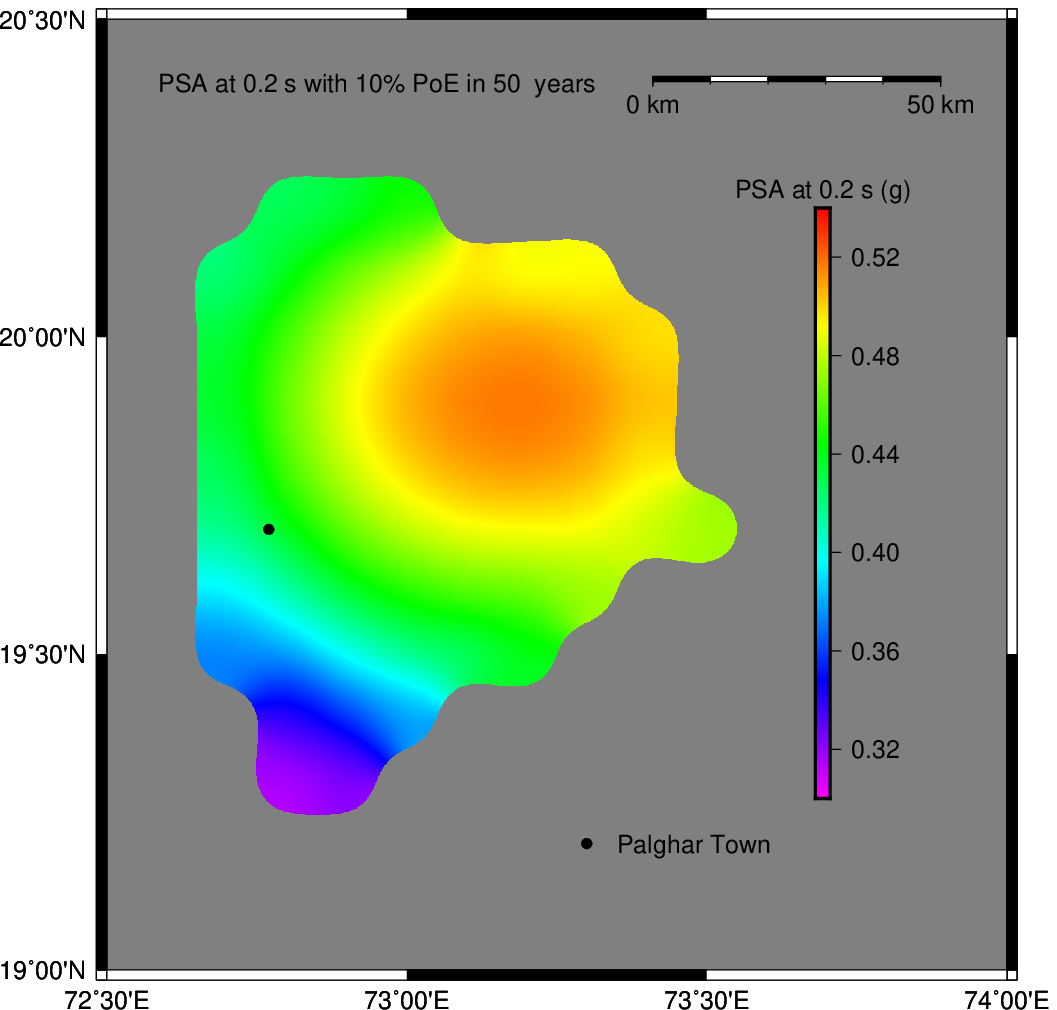}}} &
      \resizebox{65mm}{!}{\rotatebox{0}{\includegraphics[scale=0.6]{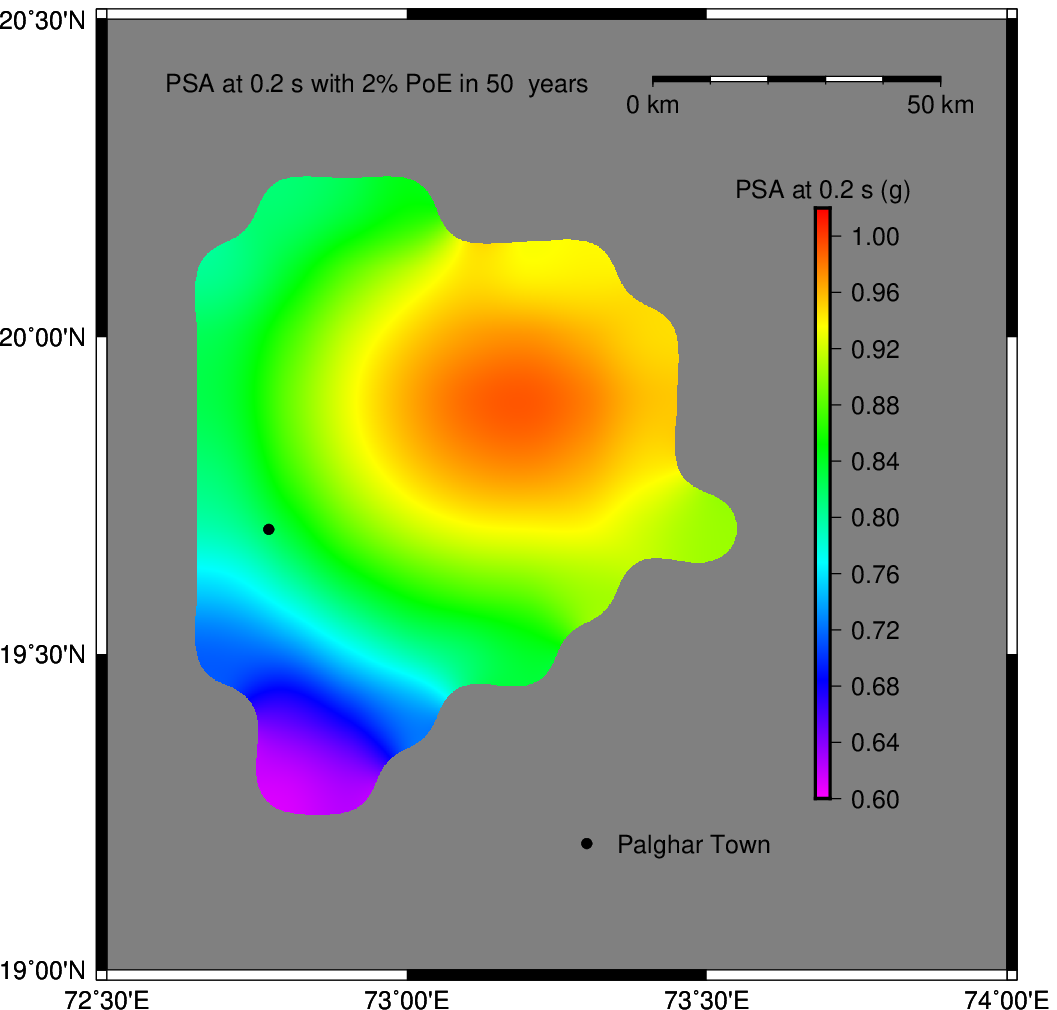}}} \\
      \resizebox{65mm}{!}{\rotatebox{0}{\includegraphics[scale=0.6]{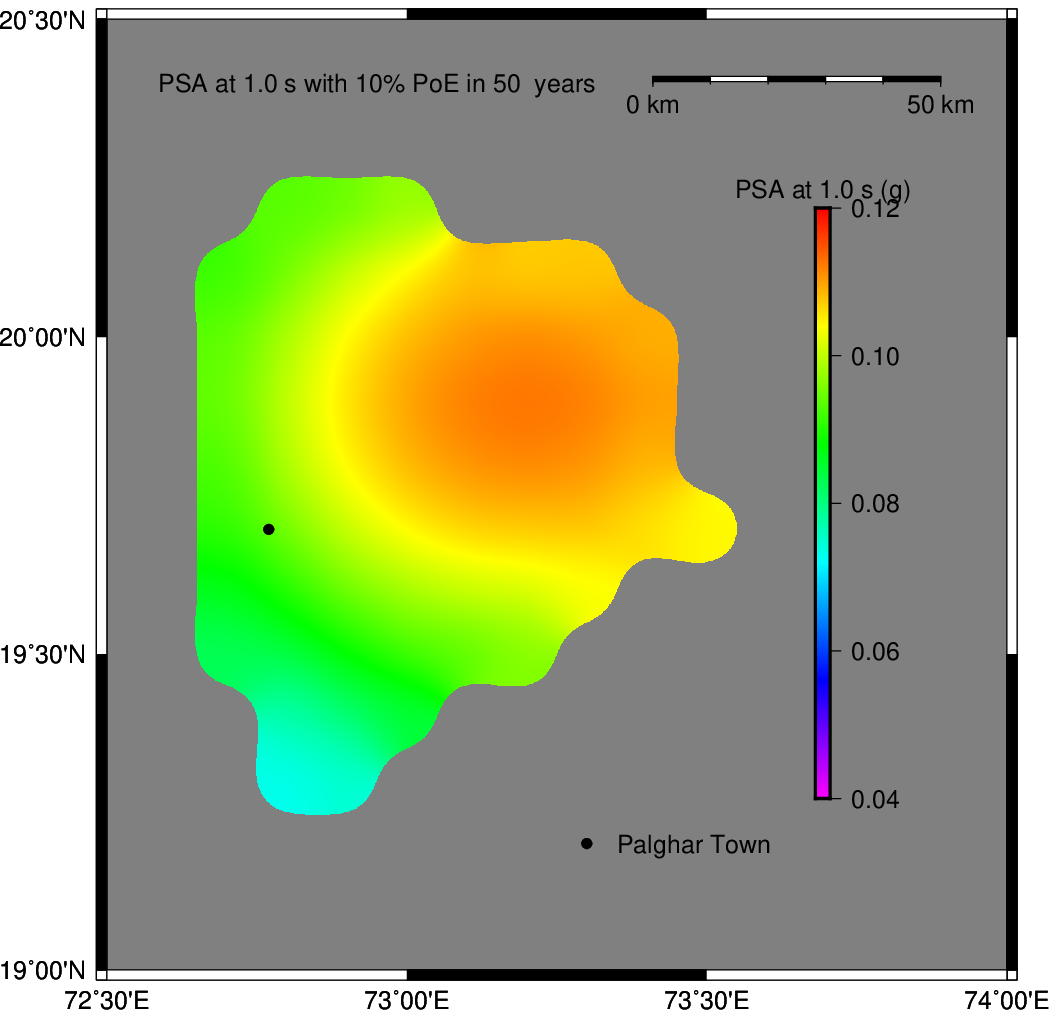}}} &
      \resizebox{65mm}{!}{\rotatebox{0}{\includegraphics[scale=0.6]{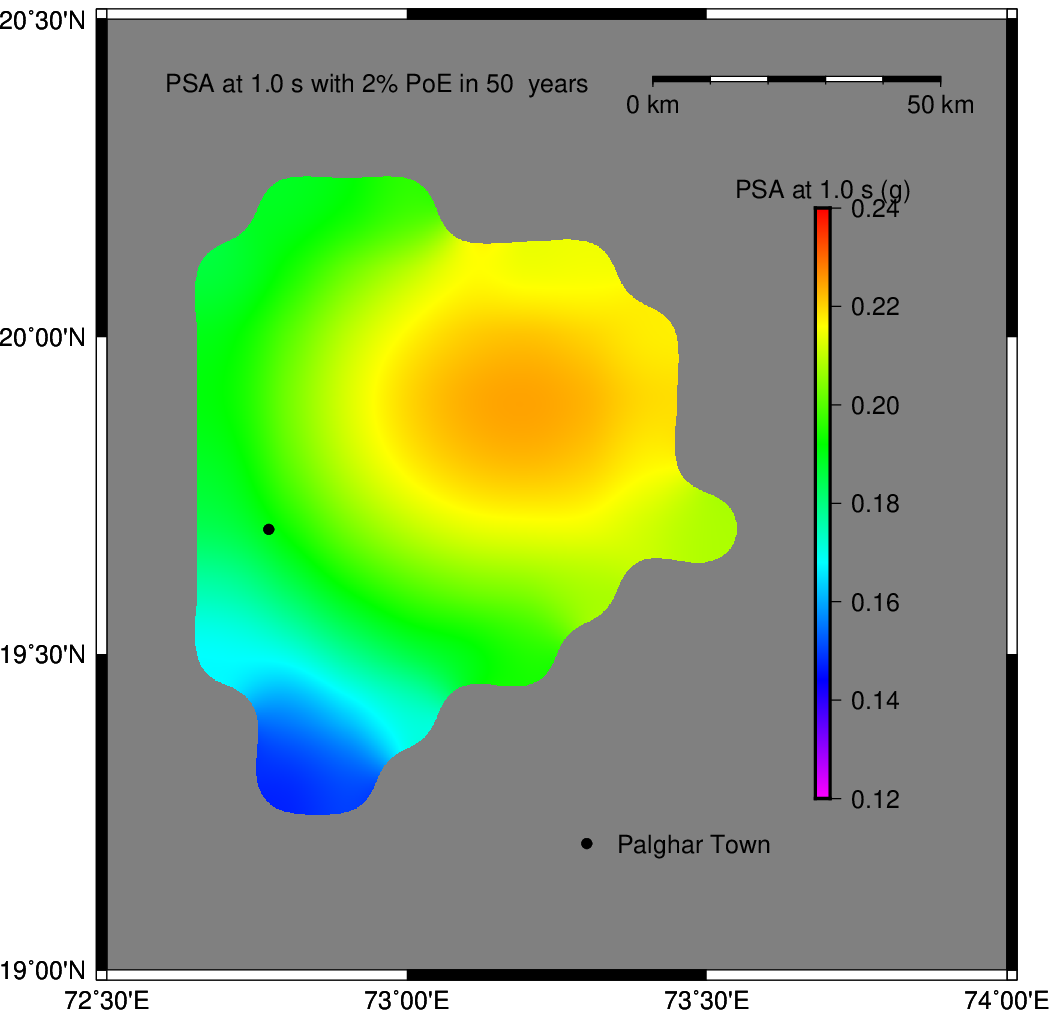}}} \\
    \end{tabular}
\end{center}
\caption{Map showing the seismic hazard distribution over Palghar district with regard to PGA
        and PSA at 0.2 s and 1.0 s for 10$\%$ PoE in 50 years (left) and
        2$\%$ PoE in 50 years (right) at engineering bed rock condition.}
\label{shzd}
\end{figure}

At the engineering bedrock level conforming to $V_{S30} = 760$ m/s., 
PGA and 5$\%$ damped PSA at 0.2 s and 1.0 s for all the grid points 
for 10$\%$ and 2$\%$ PoE for a 50-year exposure period have been computed.
The spatial distribution of the same, commonly known as
seismic hazard maps, is shown in Figure \ref{shzd}
for MCE and DBE conditions (explanied earlier). The
left panel in Fig. \ref{shzd} shows the hazard maps for 475 years return
period (DBE) while the right panel shows the same for
2475 years return period (MCE). For structural design purpose, 10$\%$ PoE in
50 years is regarded as more suitable.
The spatial variation of PGA for 10$\%$ PoE in 50 years at engineering bed rock is 
in the range from 0.1191 g to 0.1921 g for Palghar district.  The PSA at 0.2 s
shows a spatial variation between 0.2856 g to 0.4698 g while for
1.0 s, it varies from 0.0654 g to 0.1062 g. The PGA value for Palghar
town has a hazard level to the tune of 0.16 g for DBE condition. Our present
results significantly improves upon the
deterministic zonation of BIS code \cite{bis1} and captures the local
variation in seismic hazard well, whereas BIS code \cite{bis1} suggested to
consider an uniform hazard value.

The spatial variation in PGA for 2$\%$ PoE in 50 years
ranges from 0.2109 g to 0.3135 g while PSA at 0.2 s and 1.0 s
shows a variation of 0.5203 g to 0.7842 g and 0.1273 g to
0.2021 g, respectively. The local variation in seismic hazard is evident
from the Figure \ref{shzd}. The consideration of non-uniform
seismicity together with suitability test of GMPEs, based on extensive
computation of quantitative assessment, has yielded an improved
seismic hazard level for this region. The epistemic uncertainty is also taken
care of by selecting three GMPEs with weight factors determinined from the
method provided by Scherbaum {\it et al.} \cite{sche2}.

\begin{figure}[!h]
\begin{center}
        \rotatebox{0}{\includegraphics[scale=0.5]{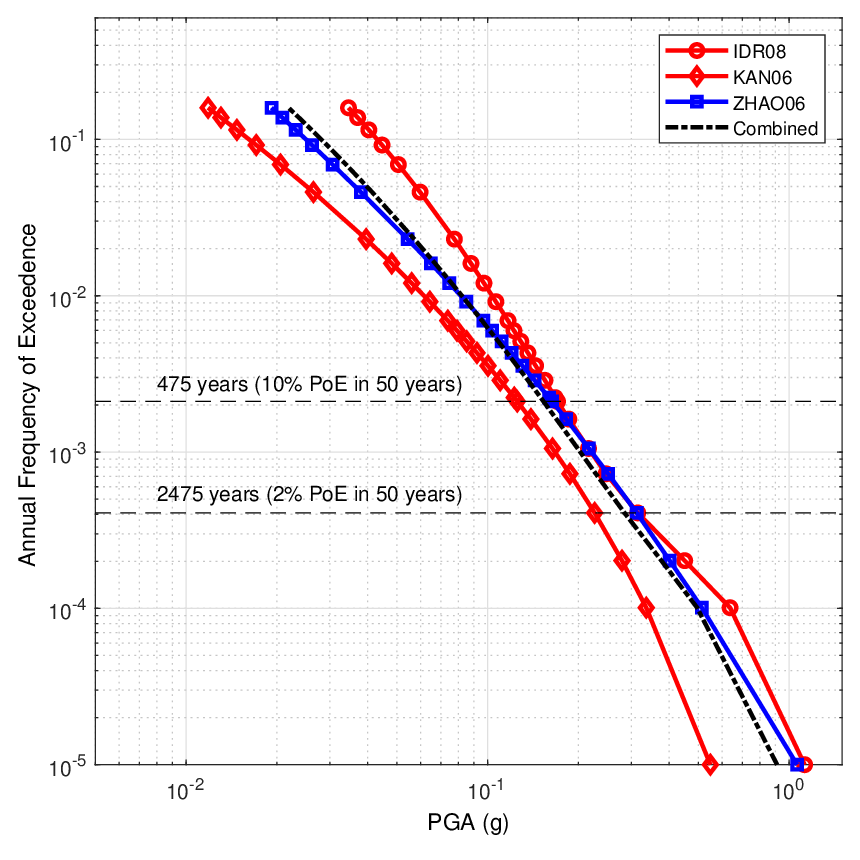}}
\caption{The hazard curves for PGA at Palghar town for different GMPEs used
        along with the combined result}
\label{hazcur}
\end{center}
\end{figure}

\begin{figure}[!h]
\begin{center}
        \rotatebox{0}{\includegraphics[scale=0.5]{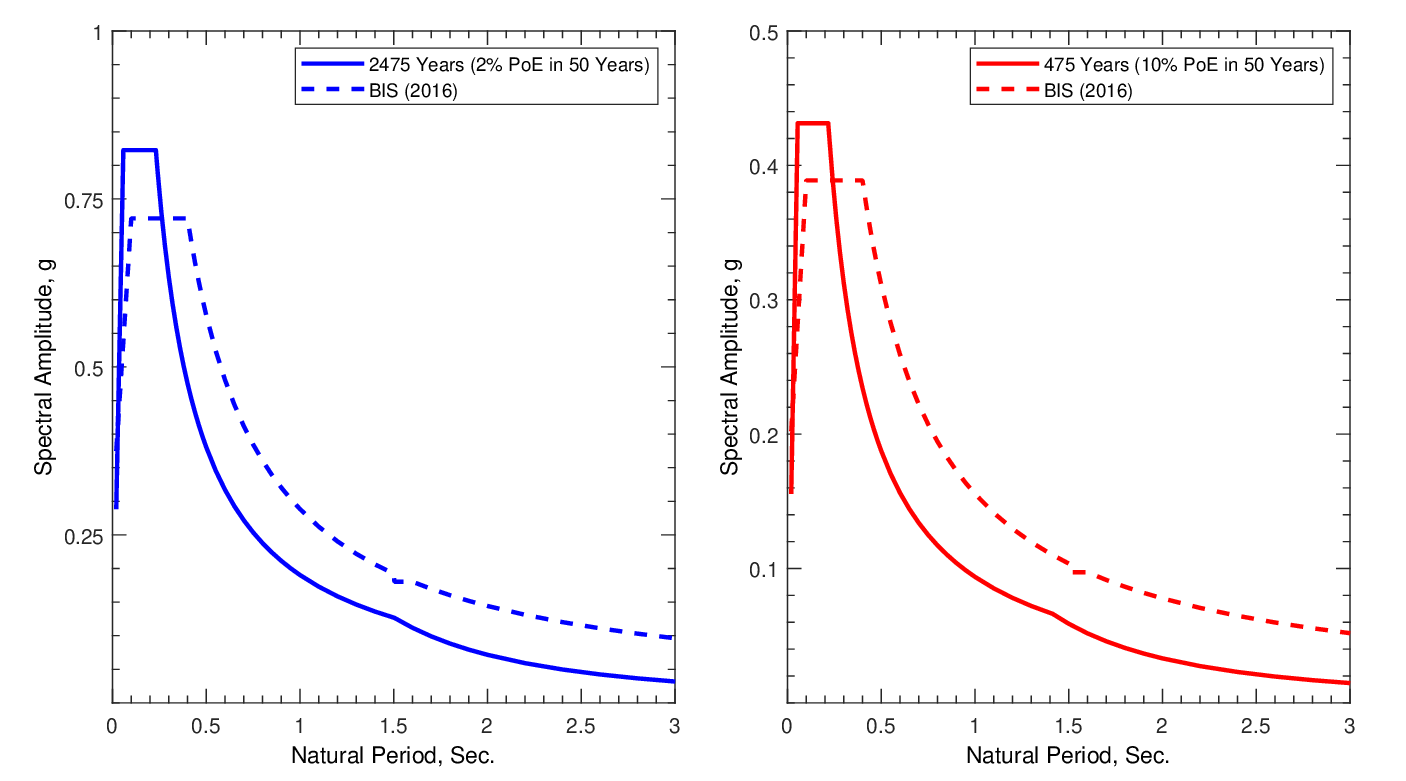}}
\caption{The 5$\%$ damped uniform hazard spectra at Palghar town compared
        with BIS code \cite{bis1}}
        \label{trs}
\end{center}
\end{figure}

As defined in subsection \ref{ct},
the seismic hazard curve (SHC) is a graphic plot showing annual frequency of exceedance (AFE)
against PGA or one of the PSAs.
The SHC for the PGA at Palghar town for the three GMPEs selected
are shown in Fig. \ref{hazcur} together with the hazard curve obtained by
combining the GMPEs with appropriate weight factors. The AFE values of 0.0021
and 0.000404 correspond to 10$\%$ and 2$\%$ probabilities of exceedance
in 50 years, equivalent to 475 years and
2475 years of return periods respectively. The SHC is also developed for PSA
at 0.2 s and 1 s at Palghar town (not shown here). The
PGA and PSA values at 0.2 s and 1 s,
determined from the SHC for the two AFE values mentioned,
are used to produce
uniform hazard spectra (UHS). The UHS
refers to a response spectrum with equal
PoE in all the time periods.
In PSHA, UHS are very important as these provide essential probabilistic
information needed for an advanced seismic hazard analysis. UHS are also
termed as target response spectra (TRS). The comparison of 5$\%$ damped UHS at
Palghar town for 10$\%$ and 2$\%$ PoE with that of BIS code \cite{bis1}
are plotted in Fig. \ref{trs}. It is noticed that the amplitudes of
spectral acceleration obtained from BIS code \cite{bis1} are on lower side compared
to those obtained from the present study for the natural periods 0.23 s and 0.22 s 
for 10$\%$ and 2$\%$  PoE respectively. The hazard curve and the
$5\%$ damped UHS for any
grid point (site) within the study area can be estimated from the hazard
maps shown in Fig. \ref{shzd}.
From the 5$\%$ damped UHS,
the acceleration time history, compatible with the UHS, and the design
response spectra for any other value of damping can easily be
computed \citep{ASCE}.
\begin{table}[!hbt]
\caption{Horizontal component of ground motion values at the location of dams
        in Palghar district and TAPS}
        \label{damsa}
\begin{adjustbox}{max width=1.1\textwidth,center}
\begin{tabular}{lcccccccc}
\hline
        Location & Lon ($\degree E$) & Lat ($\degree N$) &
        \multicolumn{3}{c}{DBE (Horizontal) (g)} &
        \multicolumn{3}{c}{MCE (Horizontal) (g)} \\
        & & & PGA & PSA (0.2 s) & PSA (1.0 s) & PGA & PSA (0.2 s) & PSA (1.0 s) \\
\hline
TAPS	&	72.6617	&	19.8294	&	0.15192	&	0.41708	&	0.09074	&	0.28165	&	0.79489	&	0.18506	\\
Dhamini	&	73.0608	&	19.9251	&	0.18343	&	0.51301	&	0.11112	&	0.33800	&	0.96193	&	0.21981	\\
Kurze	&	72.9493	&	20.0757	&	0.17054 &	0.47499	&	0.10297	&	0.32581	&	0.93751	&	0.21299	\\
Vandri	&	72.9539	&	19.6149	&	0.16683	&	0.46384	&	0.10089	&	0.29756	&	0.84543	&	0.19648	\\
Vaitarna	&	73.2906	&	19.6708	&	0.17357	&	0.48401	&	0.10596	&	0.32521	&	0.93482	&	0.21353	\\
Palgarh Town	&	72.7699	&	19.6967	&	0.15648	&	0.43442	&	0.09438	&	0.29365	&	0.84054	&	0.19350	\\
\hline
\end{tabular}
\end{adjustbox}
\end{table}
The significance of the study lies in calculating the level of
seismic hazard in a realistic way by considering the most
suitable GMPEs in a region that hosts a number of dams and
a nuclear power station. Since selection of GMPEs plays the
most crucial role in PSHA, the hazard maps obtained from
a combination of suitable GMPEs for this region makes the
study relevant in designing new earhquake-resistant structures
and evaluating seismic safety of existing structues. The outcome
of the study will help in seismic microzonation for urban planning
and land use management.
\section{Conclusion}
\label{conclu}
The purpose of the present research is to get the probabilistic
seismic hazard map of Palghar district, Maharashtra, as the 
Palghar district has witnessed an unusual frequency of earthquakes recently.
As the district houses an atomic power station and a number of dams,
it is of extreme economical importance. Therefore, re-evaluation of
seismic safety in terms of seismic hazard in a practical way for
this region is very necessary. As the ground motion parameters are
highly reliant on the choice of the GMPEs, it is indeed necessary
to give due importance on the selection of appropriate GMPEs. The
selection of GMPEs for a particular region from established methodology helps us
to overcome the subjective judgement of choosing GMPEs. The present
work has addressed the issue of selecting suitable GMPEs in details
and attempted to improve upon the previous works to some extent.
An updated and
comprehensive earthquake catalogue, homogenized and declustered
subsequently, has been employed in the present
research.
As the seismicity in Peninsular India is diffused in nature,
smooth-gridded seismicity model is more appropriate than uniform areal
seismicity model. The utilization of several GMPEs with appropriate weight factors,
selected through detailed quantitative assessment
using recorded strong motion data,
accounts for the epistemic uncertainties, thus improving the results of
the present study. The hazard computations have been performed in
a finer grid resolution of 0.02$\degree$ $\times$ 0.02$\degree$,
which results in more accurate values of PGA and PSA.
The probabilistic
seismic hazard maps of Palghar district in engineering bed rock condition
with regard to PGA and 5$\%$ damped PSA at 0.2 s and 1.0 s for
475 years and 2475 years return periods are presented. 
The hazard maps show considerable improvements
over the previous studies, which is reflected in
the spatial variation of the PGA and the PSA.
The hazard curves and the UHS are also computed for
Palghar town and the comparison of UHS with the BIS code \cite{bis1} 
are shown. As PSHA is accepted to be the most rational means to
quantify the seismic hazard \citep{col1}, the present research focuses
on developing seismic hazard maps in terms of PGA and PSA by following
an advanced seismic design philosophy. Since a structure is expected
to face all possibilities of occurrence of ground motion in its
design life, the results, obtained from this PSHA study, can be used for
designing and constructing earthquake resistant
structures in the area of the study, in addition to assessing seismic
safety of the existing structures and serving for urban planning by
identifying areas having different seismic hazard potential, known as
microzonation.
\section*{Acknowledgements}

Prof. A. Kijko is gratefully acknowledged by one of the authors, S. Sinha, 
for providing the computer program that determines $M_{\tt max}$. The Generic 
Mapping Tools (GMT) software package \citep{wess}was utilized to prepare 
Figures \ref{ros}, \ref{sgs} and \ref{shzd}. The authors express their sincere gratitude 
to Y.N. Srivastava, Additional Director, CWPRS, for his continuous encouragement.
\section*{Authors' Statements}
Suman Sinha and S. Selvan were involved in conceptualization,
analysis and writing of the manuscript.
Collection of data and material preparation were carried out by
Suman Sinha, S. Selvan and Sachin Khupat. All the authors have
reviewed the manuscript.
\section*{Disclosure statement}
Regarding the content of this article, the authors have no conflicts of interest to disclose.
\section*{Funding details}
No funding was received for conducting this study.

\bibliographystyle{ama}
\bibliography{refs_EER}

\end{document}